\documentclass[superscriptaddress, prd, aps,amsmath,amssymb,showpacs,showkeys, onecolumn]{revtex4}
\usepackage[dvips]{graphicx,color}
\usepackage{times}
\usepackage{placeins}
\usepackage{xcolor}
\usepackage{caption}
\usepackage{soul,xcolor}
\usepackage[%
colorlinks=true,
urlcolor=blue,
linkcolor=red,
citecolor=blue
]{hyperref}

\begin{document}
	\title{Quasinormal Modes and Greybody factors of {\color{black}de Sitter} Black holes surrounded by Quintessence in Rastall gravity}
	
	\author{Dhruba Jyoti Gogoi \footnote{Corresponding author}}
	\email[Email: ]{moloydhruba@yahoo.in}
	\affiliation{Department of Physics, Moran College, Moranhat, Charaideo 785670, Assam, India.}
	\affiliation{Theoretical Physics Division, Centre for Atmospheric Studies, Dibrugarh University, Dibrugarh
		786004, Assam, India.}
	\author{N. Heidari}
	\email[Email: ]{heidari.n@gmail.com}
	\affiliation{Physics Department, Shahrood University of Technology, Shahrood, Iran.}

	\author{J. K\u{r}\'\i\u{z} }
	\email[Email: ]{jan.kriz@uhk.cz}
	\affiliation{Department of Physics, University of Hradec Kr$\acute{a}$lov$\acute{e}$, Rokitansk$\acute{e}$ho 62, 500 03 Hradec Kr$\acute{a}$lov$\acute{e}$, Czechia.}
	
	\author{H. Hassanabadi}
	\email[Email: ]{hha1349@gmail.com}
	\affiliation{Physics Department, Shahrood University of Technology, Shahrood, Iran.}
	\affiliation{Department of Physics, University of Hradec Kr$\acute{a}$lov$\acute{e}$, Rokitansk$\acute{e}$ho 62, 500 03 Hradec Kr$\acute{a}$lov$\acute{e}$, Czechia.}

	\begin{abstract}
		
		In this work, we have studied the quasinormal mode, greybody factors, and absorption cross section of {\color{black}de Sitter} Reissner-Nordstr\"om black hole surrounded by quintessence field in Rastall gravity. The violation of energy-momentum conservation has a non-linear effect on the quasinormal modes. 
		With an increase in the black hole charge, both real parts of quasinormal modes i.e. oscillation frequency of ring-down Gravitational Waves (GWs) and damping or decay rate of GWs increase non-linearly. A similar observation is made for the black hole structural parameter also, however in this case the variation is almost linear. {\color{black} The remnant mass of a black hole depends on different physical parameters of a black hole.} In the case of greybody factors also, we observed that both parameters have similar impacts. With an increase in these parameters, greybody factors decrease. {\color{black} Moreover, the null geodesics and the impact of Rastall gravity on the light trajectory are also investigated.}
		Our study suggests that the presence of a surrounding quintessence field may shadow the existence of black hole charges in such black hole configurations. 
		
	\end{abstract}
	
	\maketitle
	\section{Introduction}
	
	General Relativity (GR) is a cornerstone of modern physics, making two major predictions that have revolutionized our understanding of the universe: black holes and GWs. The recent groundbreaking detections of binary black hole systems by the Laser Interferometer Gravitational Wave Observatory (LIGO) and the Variability of Solar Irradiance and Gravity Oscillations (Virgo) detector systems have provided compelling evidence supporting the validity of GR. These direct observations of GWs have not only confirmed the existence of black holes but also opened up new avenues for testing different theories of gravity, including GR.
	
	GR has already withstood rigorous experimental tests in the weak field and moderately relativistic regimes, such as solar system tests \cite{will2014} and binary pulsars \cite{hulse1975, damour1992}. However, the detection of GWs by LIGO and Virgo has established the viability of GR even in the highly relativistic strong gravity regime associated with binary black holes. It is in this regime that the properties of GWs can reveal potential deviations from GR, as expected in modified theories of gravity.
	
	In various modified gravity frameworks, such as f(R) gravity, the characteristics of GWs can undergo significant changes. For instance, in the metric formalism of f(R) gravity, the polarization modes of GWs increase to three: the first two modes correspond to the tensor plus and cross modes of GR, while the third mode is a scalar polarization mode, which is a combination of a massless breathing mode and a massive longitudinal mode \cite{Liang_2017, gogoi1, gogoi2}. The tensor modes are transverse, traceless, and massless, propagating at the speed of light through spacetime. On the other hand, the massless breathing mode is transverse but not traceless. These altered properties of GWs in different Modified Theories of Gravity (MTGs) necessitate a comprehensive examination of the behaviors and properties of black holes and compact stars within the realm of MTGs.
	
	Black holes, known as some of the cleanest objects in the universe, and neutron stars represent potential sources for the generation of GWs. Recognizing this significance, extensive research efforts have been dedicated to understanding the behavior and properties of black holes and compact stars in MTGs. Investigating these astrophysical objects in the context of MTGs provides valuable insights into the fundamental aspects of gravity and the potential deviations from GR in extreme gravity regimes.
	
	Rastall gravity, an alternative extension of GR, has gained considerable attention from researchers in recent years. Although originally introduced by P. Rastall in 1972, it did not receive significant recognition at the time. However, due to its unique characteristics, including the violation of the normal conservation law in the presence of non-vanishing background curvature, Rastall gravity has garnered increased interest. In this modified theory, the original conservation law is modified by establishing a proportional relationship between the covariant divergence of the stress-energy tensor and the covariant divergence of the Ricci curvature scalar. Notably, the usual conservation law can be recovered by setting the background curvature to zero, indicating that Rastall gravity is equivalent to GR in the absence of any matter source. While Rastall gravity has been shown to exhibit some equivalence to GR in certain aspects \cite{visser2018}, it becomes evident that significant deviations from GR can arise when non-zero curvature or dark energy fields are present in the theory \cite{darabi2018}.
	
	In this study, our focus lies in examining the scalar quasinormal modes for black holes within the framework of Rastall gravity. Quasinormal modes represent complex numbers associated with the emission of GWs from perturbed compact and massive objects in the universe. The real part of these quasinormal modes relates to the emission frequency, while the imaginary part corresponds to its damping \cite{Vishveshwara, Press, Chandrasekhar, Ferrari, Kono2003}.

	{\color{black} Investigations in the framework of Rastall gravity have recently offered fascinating insights into the behavior of black holes and neutron stars within this relatively new paradigm. Oliveira et al. (2015) conducted a remarkable study on neutron stars in the framework of Rastall gravity, offering insights into the impact of this modified theory on these astronomical phenomena \cite{Oliveira}. Furthermore, recent research on GW echoes from compact stars within the realm of Rastall gravity has revealed the profound influence of energy-momentum conservation violations on the structure and observable features of compact stars \cite{gogoi202303}. 
		
		In 2017, Heydarzade et al. explored the intriguing realm of black hole solutions in Rastall gravity \cite{Heydarzade}, while the same year, Heydarzade and Darabi investigated diverse black hole solutions coupled with a perfect fluid in the context of Rastall gravity \cite{Heydarzade2}.
		
		Furthermore, the investigation of quasinormal modes for black holes within the framework of GR, in conjunction with a quintessence field, has received considerable attention and has been extensively investigated in the literature \cite{Chen, Zhang, Zhang2, Ma, Zhang3, NonlinearRastall, Badawi2020, snew, Shao2020, Cai2020, HuYu2020, Shao2022}. Furthermore, a further line of research has focused on the quasinormal modes of higher-dimensional black holes enclosed by a quintessence field inside Rastall gravity, exhibiting fascinating departures from the predictions of GR \cite{Graca}.
		
		In this context, Jun Liang's comprehensive study on the quasinormal modes of black holes surrounded by a quintessence field in Rastall gravity \cite{Liang} offers valuable insights. The research delves into the variation of quasinormal modes concerning the Rastall parameter, unveiling that when $\kappa \lambda < 0$, the gravitational field, electromagnetic field, and massless scalar field exhibit more rapid damping, resulting in larger real frequencies compared to GR. Conversely, for $\kappa \lambda > 0$, a slower damping of these fields is observed, leading to smaller real frequencies of oscillations. This study also indicates that the variation patterns of the real and imaginary frequencies of quasinormal modes concerning $\kappa \lambda$ remain similar for different values of $l$ and $n$. Notably, the study by Liang focused on a fixed type of surrounding field and structural parameter $N_s$, while assuming charge-neutral black holes. Several other noteworthy studies in this field have contributed to our understanding of the implications of Rastall gravity, including Ref.s \cite{Xu, Lin, Hu, NonlinearRastall, gogoi4}.
		
	}

	Apart from Rastall gravity, quasinormal modes of black holes in different gravity theories have been investigated \cite{gogoi202301,gogoi202302,gogoi202304,gogoi202305,gogoi202306,gogoi202307,gogoi202308,n5,n6}. A comparison of such studies with this study can shed some more insights into the behavior of the quasinormal mode spectrum which can be utilized to differentiate between different gravity theories in the near future with observational results of quasinormal modes.
	{\color{black}
		Moreover, exploring the thermodynamics of the black hole in the Rastall gravity framework could clarify significant aspects of black hole physics like remnant mass and evaporation process \cite{revision, Lobo, Ali, chakra}. Thermodynamic properties of a Schwarzchild-like black hole surrounded by a perfect fluid and in quintessence field in Rastall gravity has been studied in Ref.\cite{Lobo} has discussed  Gravity Effects of Rastall gravity on Hawking Radiation from Charged Black Strings has been investigated in Ref. \cite{Ali}. The universal thermodynamics in Rastall gravity has been analyzed in \cite{chakra}.} Recently, in Ref. \cite{dj202301}, an anti-de Sitter black hole solution in Rastall gravity surrounded by quintessence field in the presence of linear charge distribution has been considered  {\color{black} which} shows that the violation of energy-momentum conservation has notable consequences on the black hole the Hawking temperature.\\
	{\color{black}Furthermore, it would be of great importance to discuss the impact of Rastall gravity on the nature of geodesic deviation and the shadow radius of different black holes \cite{Qian, abbas,bezerra} as the direct image of the shadow of M87* black hole by the Event Horizon Telescope collaboration \cite{M871, M872} provides an exciting opportunity to examine the modified theories of gravity.}
	
	Building upon the existing research, this study aims to further investigate the quasinormal modes and greybody factors of the black hole solution investigated in Ref. \cite{dj202301},  {\color{black}the Reissner-Nordstr\" om- black holes in Rastall gravity surrounded by a quintessence field in the presence of a non-vanishing cosmological constant (de-Sitter)}. By exploring the scalar quasinormal modes, we aim to deepen our understanding of the effects and implications of Rastall gravity on the dynamics and characteristics of black holes.
	The following sections will present our methodology, analysis, and findings regarding {\color{black}the thermodynamic and impact of Rastall gravity on remnant mass}, the scalar quasinormal modes. On the other hand, the greybody factors of the specific black hole solution in Rastall gravity will be found. {\color{black} The absorption cross section will be calculated and the effects of different parameters will be investigated. Moreover, the null geodesics and the effect of the Rastall gravity parameter on the light trajectory will be examined}
	
	The paper is organized as follows: In Section \ref{sec2}, we have discussed the field equations in Rastall gravity surrounded by quintessence field and the dS black hole solution associated with it. {\color{black}The metric and associated horizons are examined in Section \ref{sec3}. Thermodynamics and the remnant mass of the black hole are investigated in Section \ref{sec4}.} In Section \ref{sec5}, we have discussed scalar perturbation and effective potential behaviors. Section \ref{sec6} deals with the quasinormal modes and WKB approximation method. In Section \ref{sec7}, we have discussed the time evolution of scalar perturbation. Section \ref{sec8} discusses the greybody factor associated with the black hole. {\color{black}Section \ref{sec9} devotes to investigation of geodesics of the black hole.} Finally, in Section \ref{sec10}, we provide a brief concluding remark based on the findings.
	
	Throughout the manuscript, we use $G=c=1.$
	
	\section{Field equations and Black hole solution} \label{sec2}
	Rastall gravity introduces modifications to the standard GR by disregarding 
	the covariant conservation condition $T^{\mu \nu}_{\ ; \, \nu} = 0$. Instead, 
	a more generalized conservation condition is adopted, expressed as
	
	\begin{equation}\label{r1}
		\nabla_\nu T^{\mu\nu} = a^\mu.  
	\end{equation}
	
	To make the theory consistent with GR, $a^\mu$ is defined as,
	\begin{equation}\label{r2}
		a^\mu = \lambda \nabla^\mu R. 
	\end{equation}
	Here $\lambda$ is a model parameter responsible for the violation of energy-momentum conservation. Utilising Eq.s \eqref{r1} and \eqref{r2}), one can obtain the field equations for the Rastall gravity as,
	\begin{equation}
		R_{\mu\nu}-\frac{1}{2}\left(\, 1 -2\beta\,\right) g_{\mu\nu}R=\kappa T_{\mu\nu}\ ,\label{E0}
	\end{equation}
	where the term $\beta = \kappa \lambda$ is known as the Rastall parameter. Trace of the above equation provides
	\begin{equation}
		R=\frac{\kappa}{\left(4\,\beta-1\right)}\,T\ ,\quad \beta\ne 1/4. \label{E2}
	\end{equation}
	
	In the presence of a non-vanishing cosmological constant $\Lambda$, we can rewrite the field equations as
	\begin{equation}\label{E1}
		G_{\mu\nu}+ \Lambda g_{\mu \nu} + \beta g_{\mu\nu}  R  = \kappa T_{\mu\nu},
	\end{equation}
	where $G_{\mu\nu}$ stands for the standard Einstein tensor.
	
	Now, considering a spherically symmetric spacetime metric ansatz,
	\begin{equation} \label{metric}
		ds^2 = -f(r) dt^2 + \dfrac{dr^2}{f(r)} +r^2 d\Omega^2,
	\end{equation}
	and defining the Rastall tensor as $\Theta_{\mu\nu} = G_{\mu\nu}+\Lambda g_{\mu\nu} + \kappa \lambda g_{\mu\nu} R$, the non-vanishing components of the field equation are found to be:
	\begin{eqnarray}\label{H}
		&&{\Theta^{0}}_{0}=\frac{1}{r^2}\big[rf^{\prime}(r) + f(r) -1 \big]+ \Lambda+\beta R,\nonumber\\
		&&{\Theta^{1}}_{1}=\frac{1}{r^2}\big[rf^{\prime}(r) + f(r) - 1  \big]+ \Lambda+\beta R,\nonumber\\
		&&{\Theta^{2}}_{2}=\frac{1}{r^2}\big[rf^{\prime}(r)+\frac{1}{2}r^2 f^{\prime\prime}(r)\big]+ \Lambda+\beta R,\nonumber\\
		&&{\Theta^{3}}_{3}=\frac{1}{r^2}\big[rf^{\prime}(r)+\frac{1}{2}r^2 f^{\prime\prime}(r)\big]+ \Lambda+\beta R,
	\end{eqnarray}
	where one can have
	\begin{equation}\label{R}
		R=-\frac { 1
		}{{r}^{2}}\big[{r}^{2}f^{\prime\prime}(r) + 4r f^{\prime}(r) + 2\,f(r) -2 \big].
	\end{equation}
	The total stress-energy tensor $T^\mu_\nu$ in this case is defined by
	\begin{equation} \label{gen_T}
		{{T}^{\mu}}_{\nu}={E^{\mu}}_{\nu}+{\mathcal{T}^{\mu}}_{\nu},
	\end{equation}
	where the first term on the right hand side is the Maxwell tensor which is represented as
	\begin{equation}\label{E**}
		{E^{\mu}}_{\nu}={\frac{Q^2}{\kappa r^4}}~\begin{pmatrix}-1 & 0 & 0 & 0 \\
			0& -1 & 0 & 0 \\
			0 & 0 & 1 & 0 \\
			0 & 0 & 0 & 1\\
		\end{pmatrix}.
	\end{equation}
	
	The other term on the right hand side of Eq.\ref{gen_T} represents a surrounding dark energy field that has the following non-zero components:
	\begin{eqnarray}
		&&{\mathcal{T}^{0}}_{0}={\mathcal{T}^{1}}_{1}=-\rho_s,\nonumber\\
		&&{\mathcal{T}^{2}}_{2}={\mathcal{T}^{3}}_{3}=\frac{1}{2}(1+3\omega_s)\rho_s.
	\end{eqnarray}
	
	Finally, the contributing components ${\Theta^{0}}_{0}={T^{0}}_{0}$ and ${\Theta^{1}}_{1}={T^{1}}_{1}$ eventually will provide:
	\begin{equation}\label{e00}
		\frac{1}{r^2}\big(rf^{\prime} + f  -1 \big) + \Lambda-\frac {\beta}{{r}^{2}}\big({r}^{2}f^{\prime\prime} + 4r f^{\prime}+2\,f -2\big)=-\kappa\rho_s-\frac{Q^2}{ r^4},
	\end{equation} 
	and ${\Theta^{2}}_{2}={T^{2}}_{2}$ and ${\Theta^{3}}_{3}={T^{3}}_{3}$ components can be expressed as
	\begin{align}\label{e22}
		\frac{1}{r^2}\big(rf^{\prime}+\frac{1}{2}r^2 f^{\prime\prime}\big) &+ \Lambda-\frac {\beta
		}{{r}^{2}}\big({r}^{2}f^{\prime\prime}  +4r f^{\prime} +2\,f  -2 \big)
		\notag \\&=\frac{1}{2}(1+3\omega_{s} )\kappa\rho_{s}+\frac{Q^2}{ r^4}.
	\end{align}
	{\color{black}In the next section, we explore the metric function.

		\FloatBarrier	
		\section{The metric function and horizons properties}} \label{sec3}
	In this study, we choose the quintessence field i.e. the equation of state parameter $\omega_s = -1/3$. With this assumption, the lapse function will be found as follows
	\begin{equation}\label{f1}
		f(r) = 1 - \frac{2M}{r} +\frac{Q^2}{r^2} + N_s r^{\frac{4 \beta }{1-2 \beta }} - \frac{\Lambda  r^2}{3-12 \beta},
	\end{equation}
	and the associated energy density is \cite{dj202301}
	\begin{equation}\label{rho}
		\rho_s (r)=\frac{(2 \beta +1) (4 \beta -1) N_s r^{\frac{4 \beta }{1-2 \beta }-2}}{(1-2 \beta )^2 }.
	\end{equation}
	The integration constants $M$ and $N_s$ in this context correspond to the black hole mass and the structural characteristics of the black hole's surrounding field, respectively. 
	In our investigation, we shall use this black hole spacetime which was previously used in Ref. \cite{dj202301}. 
	
	\setstcolor{black}
	One may note from Eq. \eqref{rho} that $\rho_s (r)\geq 0$ puts a bound on the Rastall parameter: {\color{black} $\beta <-1/2 \; \text{or} \; \beta > 1/4$ for $N_s >0 \;$} However, this bound may not be very suitable to describe the theory near the GR limit or for very small deviation of the theory from GR. This can be easily overcome using {\color{black} the condition $ \; -1/2 < \beta < 1/4 \;$ \text{and}\; $N_s <0$ which is applied in this study.} 
	Moreover, to keep our parameter space consistent with the observational constraints, we keep $\beta$ in the order of constraints put by Ref. \cite{Tang}.
	
	{\color{black}
		To find the horizon radius, $f\left(r\right)$ in Eq. \eqref{f1} is assumed to be zero but there is no an analytical result for it. We should consider some constant parameters and find the horizon numerically. the lapse function is plotted in Fig. \ref{fig:fr} for $M=1$, $Q=0.1$, $N_s=-0.1$, $\Lambda=0.02$ and $\beta=0.02$ and in this case,
		the metric
		implies three horizons.
		the first root is the Cauchy horizon radius, the second is the event horizon radius and the third one is the cosmological horizon radius. 
		For some specific values, the horizon disappears, therefore, we consider the variable which results in an acceptable horizon radius in this work.
		
		\begin{figure}
			\centering
			\includegraphics[scale = 0.5]{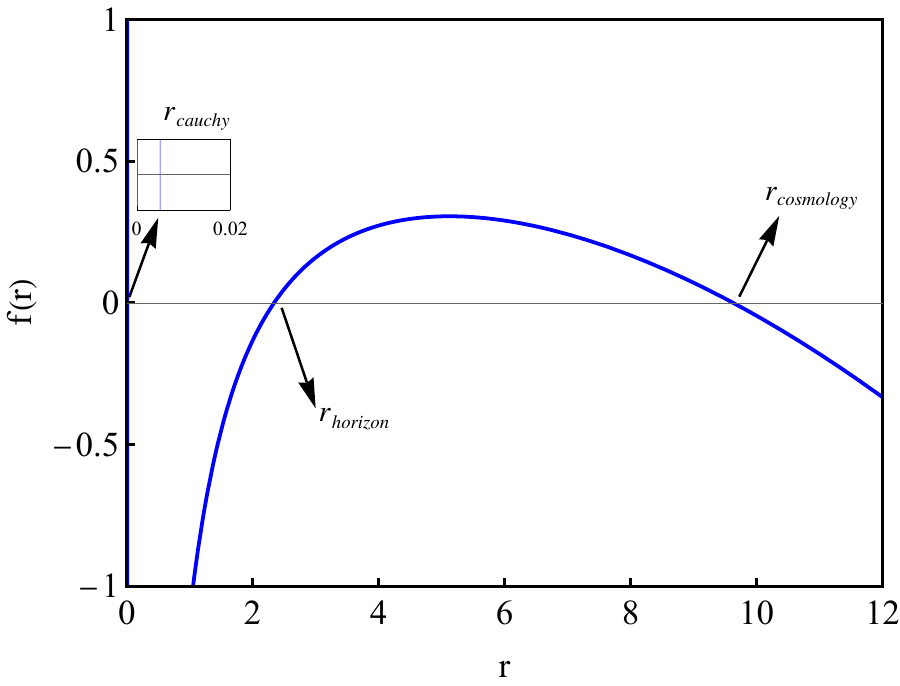}
			\caption{{\color{black}Lapse function $M=1$, $N_s= -0.1$ , $Q = 0.1$, $\beta = 0.02$ and $\Lambda = 0.02$. }}
			\label{fig:fr}
		\end{figure}	
		
		According to the lapse function represented in Eq.\ref{f1}, we try to examine the Hawking temperature of the black hole.\\
	}
	
	{\color{black}
		
		\FloatBarrier
		\section{Temperature and Remnant mass} \label{sec4}
		Following the Hawking temperature equation, by substituting the mass as a function of horizon radius,  we can obtain the Hawking temperature as a function of horizon radius $\left({r_h}\right)$.
		\begin{equation}\label{TH}
			\ {T_H} = \frac{1}{{4\pi}}{\left. {\frac{{\mathrm{d}{f(r)}}}{{\mathrm{d}r}}} \right|_{r = {r_{h}}}} =
			\frac{(2 \beta -1) \left(\Lambda  {r_{h}}^4-(4 \beta -1) (Q-{r_{h}}) (Q+{r_{h}})\right)-(2 \beta +1) (4 \beta -1) {N_s} {r_{h}}^{\frac{2}{1-2 \beta }}}{4 \pi  \left(8 \beta ^2-6 \beta +1\right) {r_{h}}^3}
		\end{equation}
		The temperature with respect to is demonstrated in Fig.\ref{fig:temp} for  $M=1$, $N_s= -0.1$ , $Q = 0.1$, $\Lambda = 0.02$ and different values of $\beta$. 
		It is obvious that when the temperature goes to zero, its associated mass will tend to some value called remnant mass. According to this figure, when the black hole evaporates, the remnant mass value depends on the $\beta$, and for higher values of $\beta$, the remnant mass decreases. 
		\begin{figure}[h]
			\centering
			\includegraphics[scale = 0.5]{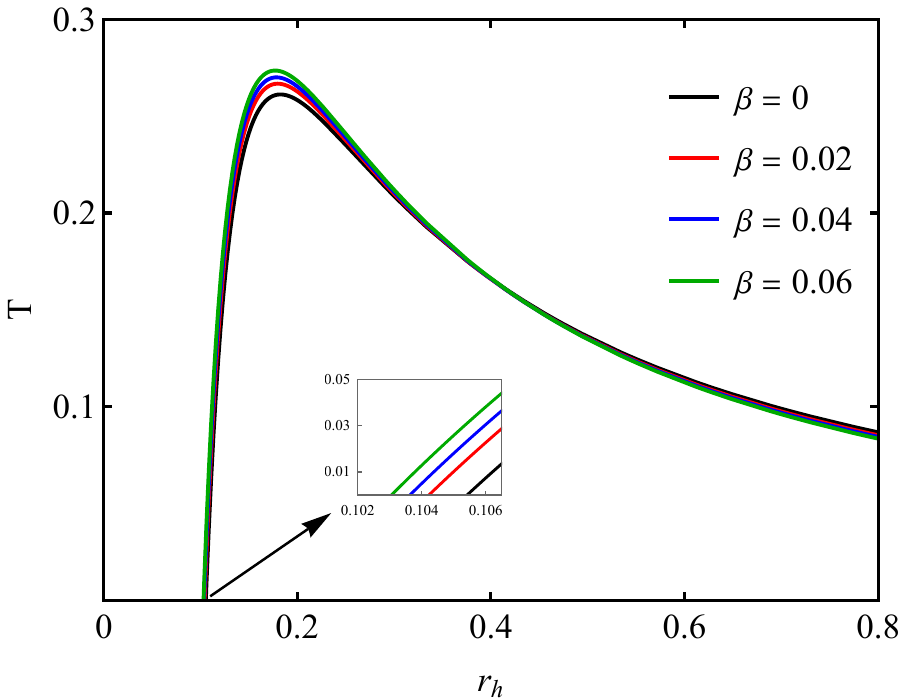}
			\caption{{\color{black} Hawking temperature as a function of horizon radius for $M=1$, $N_s= -0.1$ , $Q = 0.1$, $\Lambda = 0.02$ and different values of $\beta$.}}
			\label{fig:temp}
		\end{figure}
		Moreover, during the evaporation, black holes with a higher value of $\beta$ experience higher maximum temperature. Another aspect of the existence of maximum temperature is that through the evaporation process, the phase transition occurs. We determined the radius associated with remnant mass and the radius related to the phase transition as $r_{rem}$ and critical radius as  $r_{cr}$, respectively. The results are presented in Table. \ref{Table:remnant}. When $\beta$ increases the critical and horizon radius go down. On the other hand, the expected maximum temperature increases with the bigger value of the Rastall parameter.\\

		\begin{table}[h]
			\centering
			\caption{{\color{black} Critical radius, remnant radius, remnant mass, and maximum temperature are calculated for $M=1$, $N_s= -0.1$ , $Q = 0.1$, $\Lambda = 0.02$ and different values of $\beta$.}}
			\begin{tabular}{|c| c c c c|}
				\hline
				$\beta$ & $r_{cr}$ & $r_{rem}$ & $M_{rem}$ & $T_{Max}$ \\ \hline\hline
				0.02 & 0.181112 & 0.104829 & 0.0957637 & 0.263835 \\
				0.04 & 0.179719 & 0.104228 & 0.0965643 & 0.266792 \\
				0.06 & 0.178363 & 0.103632 & 0.0972665 & 0.270056 \\
				0.08 & 0.17708 & 0.103055 & 0.0978719 & 0.27357 \\
				0.1 & 0.175902 & 0.10251 & 0.0983837 & 0.277259 \\ \hline
			\end{tabular}
			\label{Table:remnant}
		\end{table}
		
		In the next step, we investigate the scalar field considering to obtained lapse function and its associated spacetime.
		
	}
	\section{Scalar field scattering and effective potential} \label{sec5}
	\label{sec03}
	Now, considering the metric function $f(r)$ given in Eq. \eqref{f1}, the metric described by Eq. \eqref{metric} undergoes modifications as \cite{dj202301},
	\begin{widetext}
		\begin{equation}\label{metric01}
			ds^2=-\left(1 - \frac{2M}{r} +\frac{Q^2}{r^2} + N_s r^{\frac{4 \beta }{1-2 \beta }} - \frac{\Lambda  r^2}{3-12 \beta}\right)dt^2
			+\frac{dr^2}{1 - \frac{2M}{r} +\frac{Q^2}{r^2} + N_s r^{\frac{4 \beta }{1-2 \beta }} - \frac{\Lambda  r^2}{3-12 \beta}}
			+r^2 d\Omega^2.
		\end{equation}
	\end{widetext}
	
	By assuming the modified metric in Eq. \ref{metric01}, we would like to study the scattering wave of scalar perturbation in this model. First, we
	consider a massless scalar field in this space-time, which can be expressed
	by the Klein-Gordon equation\\
	\begin{equation}\label{klein}
		\frac{1}{{\sqrt { - g} }}{\partial _\mu }(\sqrt { - g} {g^{\mu \nu }}{\partial _\nu }\Psi ) = 0
	\end{equation}	
	The spherical symmetry wave function, by decomposing variables, is assumed in the below form
	\begin{equation}\label{sai}
		{\Psi _{\omega lm}}(\mathbf{r},t) = \frac{{{R_{\omega l}}(r)}}{r}{Y_{lm}}(\theta ,\varphi ){e^{ - i\omega t}}
	\end{equation}
	where $R_{\omega l}$ is the radial part and $Y_{lm}(\theta,\phi)$ denotes the spherical harmonics. Now by applying tortoise coordinate as $dr^* =dr/f(r)$ we arrive at the following equation for radial function
	\begin{equation}\label{waves}
		\left[\frac{{{d^2}}}{{d{{r^*}^2}}} + {\omega ^2} - {{V}_{eff}}\right]R_{\omega l} (r^*) = 0
	\end{equation}
	Here the effective potential is \cite{gogoi202308}
	\begin{equation}
		V_{eff}(r) =  f(r)\left(\frac{{l(l + 1)}}{{{r^2}}}+\frac{{1}}{r}\frac{{\mathrm{d}f}}{{\mathrm{d}r}} \right)
	\end{equation}
	According to the metric in Eq. \ref{metric01} and $f(r)$ in Eq. \ref{f1}, the effective potential of scalar field are plotted for $M=1$, $\beta=0.02$ and $\Lambda=0.02$ and different values of $N_s$ and $Q$ are considered to observe the effect of these parameters on $V_{eff}$. 

\begin{figure*}[t!]
      	\centering{
      	\includegraphics[scale=0.35]{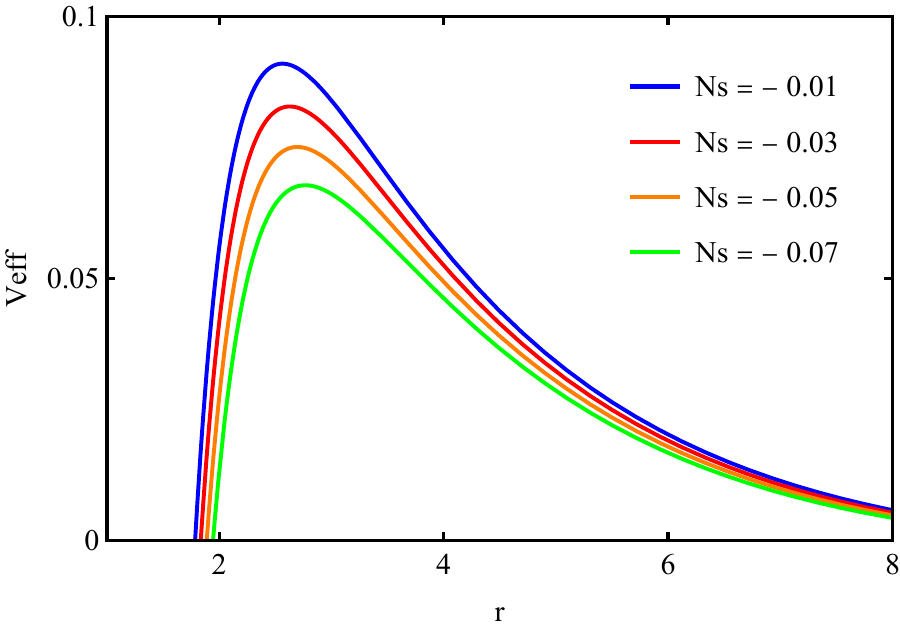}
       \includegraphics[scale=0.35]{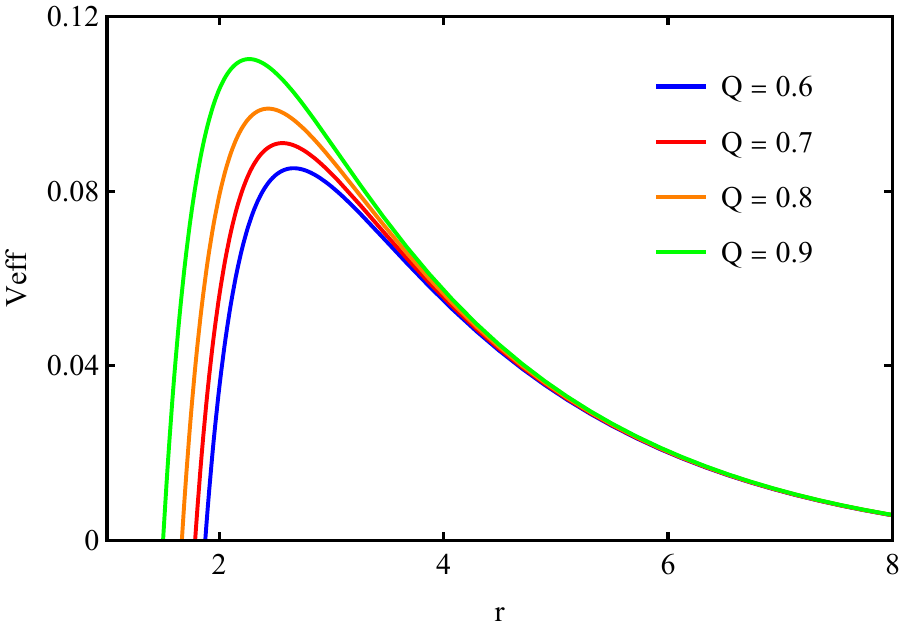}
       \includegraphics[scale=0.35]{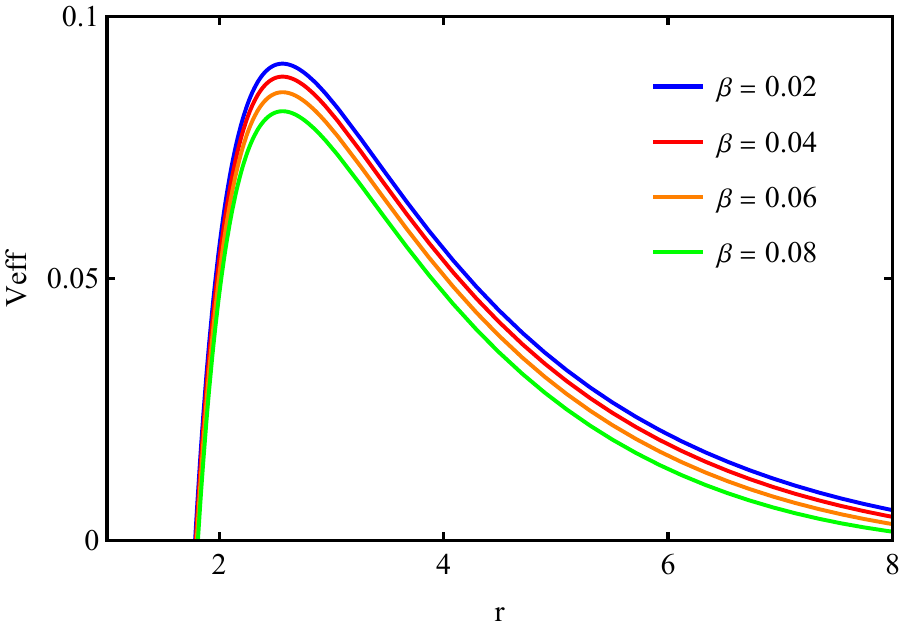}}
      	\caption{The effective potential with respect to $r$ for the scalar field and multipole $l=1$, $M=1$ and $\Lambda=0.02$. first plot is the effective potential for {\color{black} different values of $N_s$ and constant parameters $Q=0.07$ and $\beta=0.02$}, second plot is the effective potential for {\color{black} different values of $Q$, constant $N_s=-0.01$ and $\beta=0.02$} {\color{black} and third plot depicts the effective potential for different values of $\beta$, constant $N_s=-0.01$ and $Q=0.7$.}}
		\label{fig:Veff}
      \end{figure*}

	
	The Fig. \ref{fig:Veff} illustrates a notable trend caused by $N_s$, $Q$, and $\beta$ parameters as in the first panel for $N_s<0$, when $|N_s|$ increases, the maximum value of the effective potential decreases. On the other hand, the second panel shows the maximum effective potential increases by increasing $Q$ and {\color{black}the last panel demonstrates the effect of $\beta$ which leads to higher maximum effective potential for lower $\beta$.}
	As shown in Fig. \ref{fig:Veff}, parameters $Q$, $N_s$ and $\beta$ play a significant role in the shape of effective potential which affects the quasinormal modes for different values. In the next section, we would like to investigate the impact of these parameters on the quasinormal modes.
	
	\section{Quasinormal modes with WKB method}
	\label{sec6}
	
	Different methods are applied for the calculation of quasinormal modes \cite{mash, heidari, leaver}. In this section, we apply the WKB method for to obtain the quasinormal mode frequencies. In the 6th order WKB method, quasinormal modes can be evaluated with the following formula \cite{Iyer,konoplya1, Kono2003}
	\begin{equation}\label{omegawkb}
		\frac{{i(\omega _n^2 - V_{0})}}{{\sqrt { - 2V''_0} }} + \sum\limits_{j = 2}^6 {{\Omega _j} = n + \frac{1}{2}} 
	\end{equation} 
	
	Note that $V_0$ and $V''_0$ represent the height of the effective potential and the second derivative with respect to the tortoise coordinate of the potential at the maxi, respectively. The $\Omega_j$ are the correction terms which depend
	on the values of the potential and higher derivatives of it at the maxi according to \cite{Kono2003}.\\
	The results of the quasinormal modes for the scalar perturbations are given in the Tables. \ref{table:QNM1} - \ref{table:QNM2}.\\
	From Table. \ref{table:QNM1} we can see that for the constant value of $N_s$ by increasing $Q$, the real part of quasinormal modes for $l=0,1,2$ and monopoles $(n<l)$ increase which indicates the propagating frequency, increases. 
	However, the imaginary part for $l=0$ and $l=1$ or $2$  represents the damping timescale for the black hole and it does not have a particular behaviour for  $l=0,1$ and $l=2$ in this range. \\
	
	\begin{table}[!h] \label{tab01}
		\centering
		\caption{Quasinormal modes of scalar field for $l=1$, $M=1$, $\Lambda=0.02$, $\beta=0.02$, $N_s=-0.01$ and various values of $Q$ }
		\begin{tabular}{|l|l|l|l|l|}
			\hline
			~ & ~ & $Q=0.7$ & $Q=0.8$ & $Q=0.9$ \\ \hline
			$l=0$ & $n=0$ &
			$	0.1374003 - 0.144455i$ &
			$	0.1404071 - 0.143539i$ &
			$0.144138 - 0.142484i $
			\\ \hline
			$l=1$ & $n=0$ &
			
			$	0.287868\, -0.10491 i$ &
			$	0.300852\, -0.104476 i$ &
			$	0.318782\, -0.102695 i $
			\\ \hline
			$l=1$ & $n=1$ &
			
			$	0.300689 -0.311105 i $&
			$	0.312462 -0.310839 i$ &
			$	0.328737 -0.307407 i $
			
			\\ \hline
			$l=2$ & $n=0$ &	
			$			0.474914\, -0.0943035 i $&
			$	0.496649\, -0.0944301 i $ &
			$	0.52696\, -0.0933422 i $
			\\ \hline
			
			$l=2$ & $n=1$ & 
			$ 	0.473121\, -0.288294 i $&
			$ 	0.494497\, -0.288429 i $&
			$	0.524276\, -0.285016 i $
			\\ \hline
			$l=2$ & $n=2$ & 
			$0.478164\, -0.489643 i $&
			$	0.498446\, -0.489825 i $&
			$	0.526908\, -0.484446 i $ \\
			\hline
		\end{tabular}\label{table:QNM1}
	\end{table}

	In Table.\ref{table:QNM2} the quasinormal mode is represented for the constant value of $Q$ and different values of $N_s$.When the value of parameter $N_s$ decreases, the real part of quasinormal modes exhibits a decrease for $l=0,1,2$ and their monopoles. About the imaginary part of quasinormal modes when $N_s$ takes the lower value,  the imaginary part  turns out to be smaller and therefore lower damping timescale.
	
	\begin{table}[!h] \label{tab02}
		\centering
		\caption{Quasinormal modes of scalar filed for $l=1$, $M=1$, $\Lambda=0.02$, $\beta=0.02$, $Q=0.6$ and various values of $N_s$}
		\begin{tabular}{|l|l|l|l|l|}
			\hline
			~ & ~ & $N_s=-0.01$ & $N_s=-0.03$ & $N_s=-0.05$ \\ \hline
			$l=0$ & $n=0$ & $0.1349252 - 0.1448015i$ & $0.1272706 - 0.1375433i$ & $0.1197197 - 0.1303769i$ \\ \hline
			$l=1$ & $n=0$ & $0.2781200 - 0.1048082i$     & $0.2655281 - 0.0991626i$   & $0.2530085 - 0.09360868i$  \\ \hline
			$l=1$ & $n=1$ & $0.2918233 - 0.310067i$   & $0.2782748 - 0.2934625i$  & $0.2648249 - 0.2771083i$              \\ \hline
			$l=2$ & $n=0$ & $0.4586723 - 0.09379433i$ & $0.4388781 - 0.0888133i$ & $0.419147 - 0.08390988i$ \\ \hline
			$l=2$ & $n=1$ & $0.4571753 - 0.286934i$ & $0.43742 - 0.2716034 i$ & $0.4177378 - 0.25651i$ \\ \hline
			$l=2$ & $n=2$ & $0.4630524 - 0.4873898i$ & $0.4427628 - 0.4612892i$ & $0.4225712 - 0.4355807i$ \\ \hline
		\end{tabular}\label{table:QNM2}
	\end{table}

	\begin{figure}[htbp]
		\centerline{
			\includegraphics[scale = 0.5]{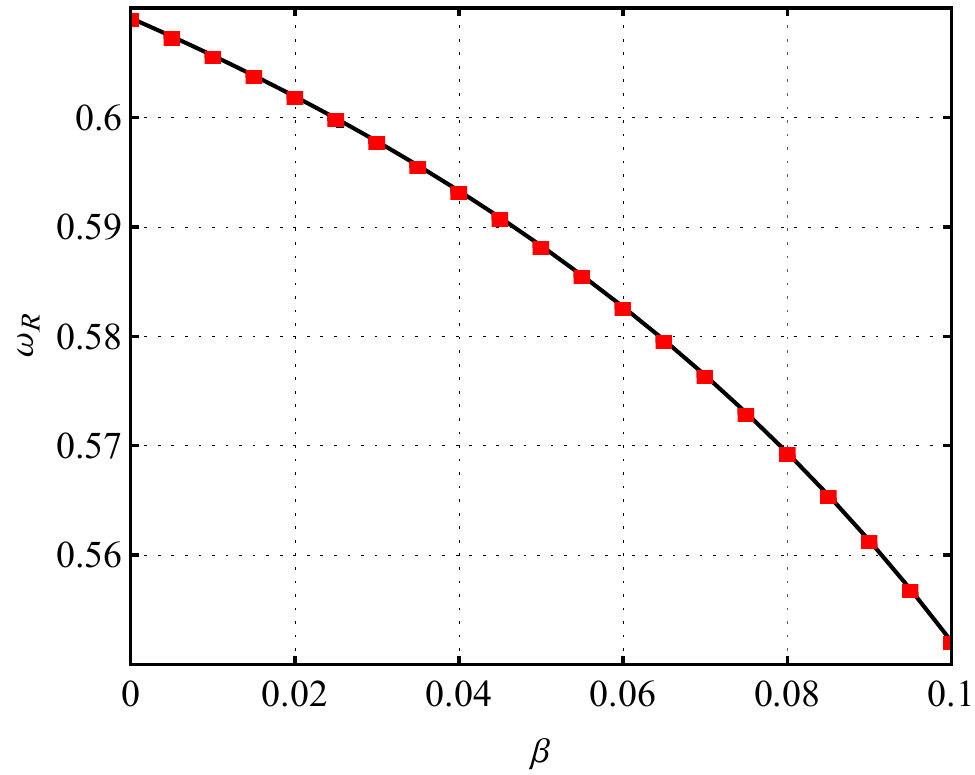}\hspace{0.5cm}
			\includegraphics[scale = 0.52]{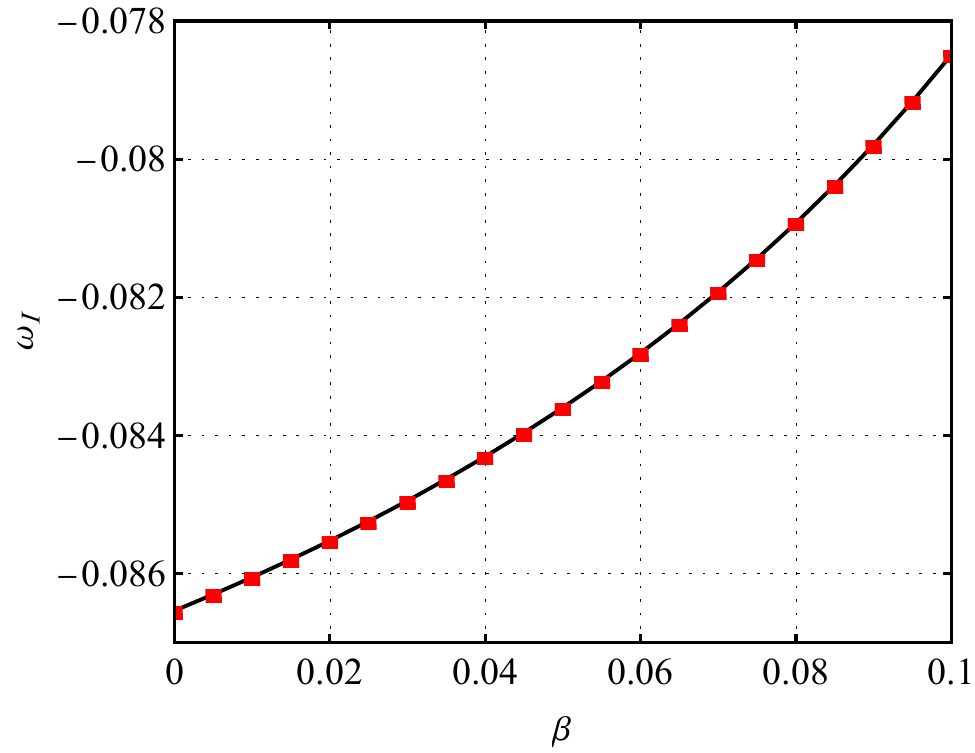}} \vspace{-0.2cm}
		\caption{{\color{black}Variation of quasinormal modes with $M=1$, $n= 0, l=3, N_s= -0.01, Q = 0.3$ and $\Lambda = 0.02$ for massless scalar
				perturbation. }}
		\label{qnm01}
	\end{figure}
	
	\begin{figure}[htbp]
		\centerline{
			\includegraphics[scale = 0.5]{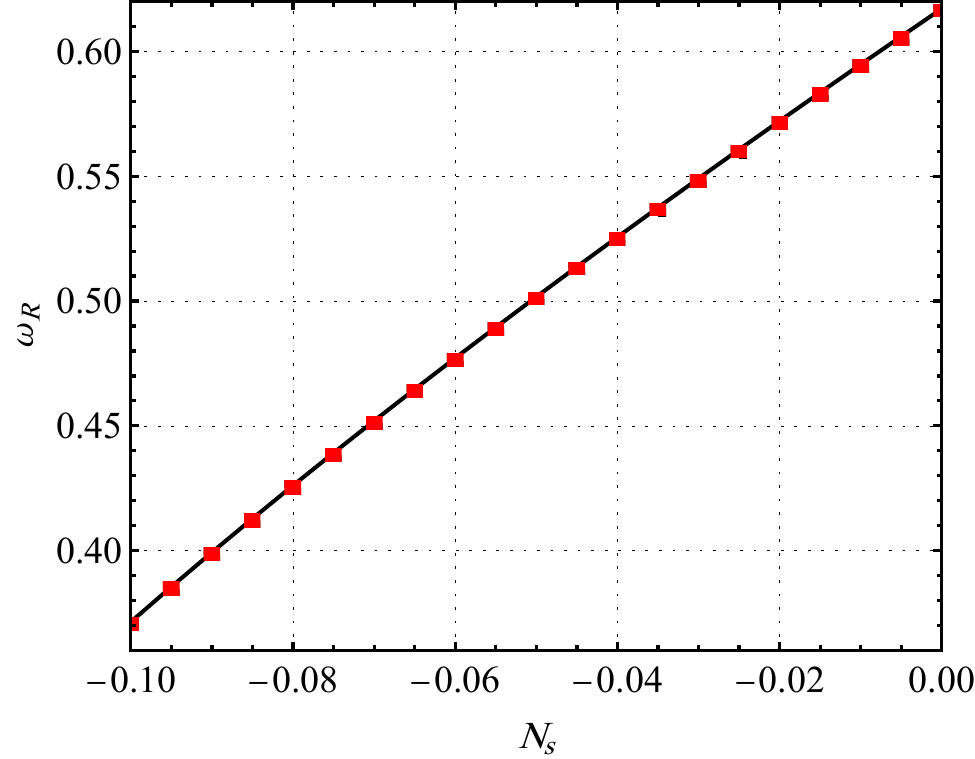}\hspace{0.5cm}
			\includegraphics[scale = 0.5]{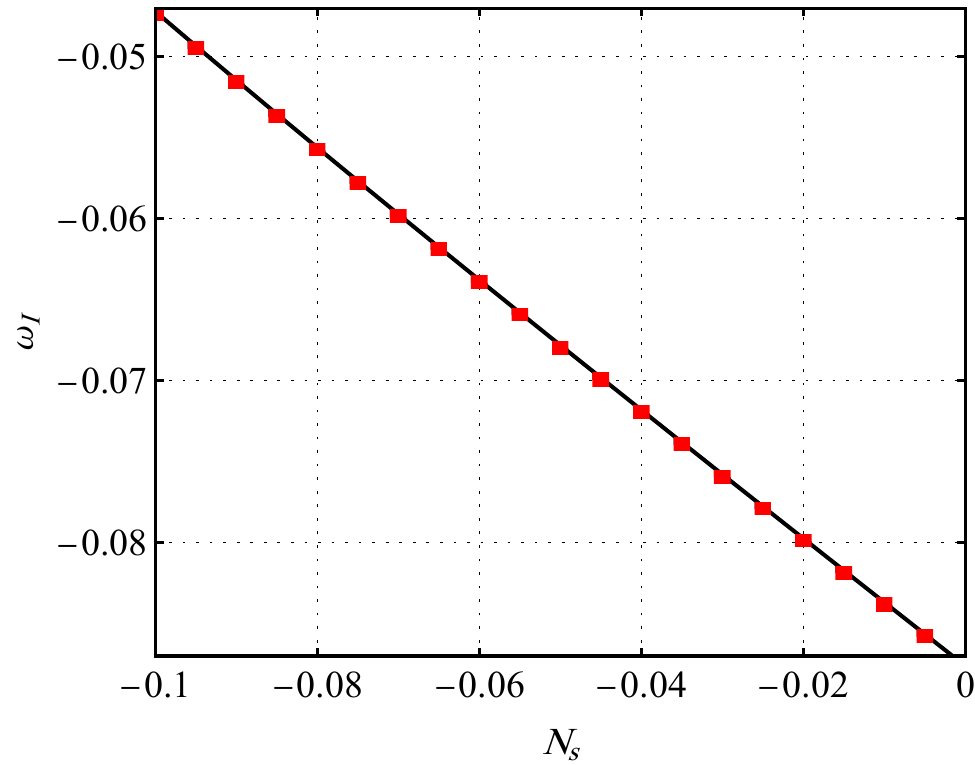}} \vspace{-0.2cm}
		\caption{\color{black}{Variation of quasinormal modes with $M=1$, $n= 0, l=3, \beta = 0.1, Q = 0.3$ and $\Lambda = 0.02$ for massless scalar
				perturbation.}}
		\label{qnm02}
	\end{figure}
	
	\begin{figure}[htbp]
		\centerline{
			\includegraphics[scale = 0.5]{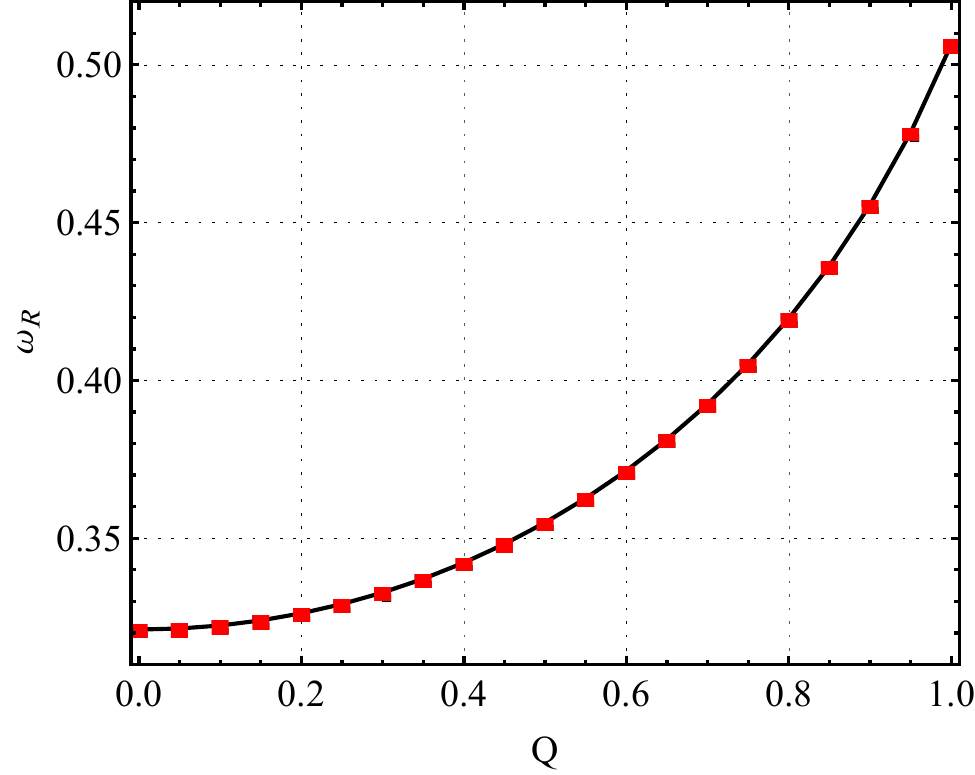}\hspace{0.5cm}
			\includegraphics[scale = 0.52]{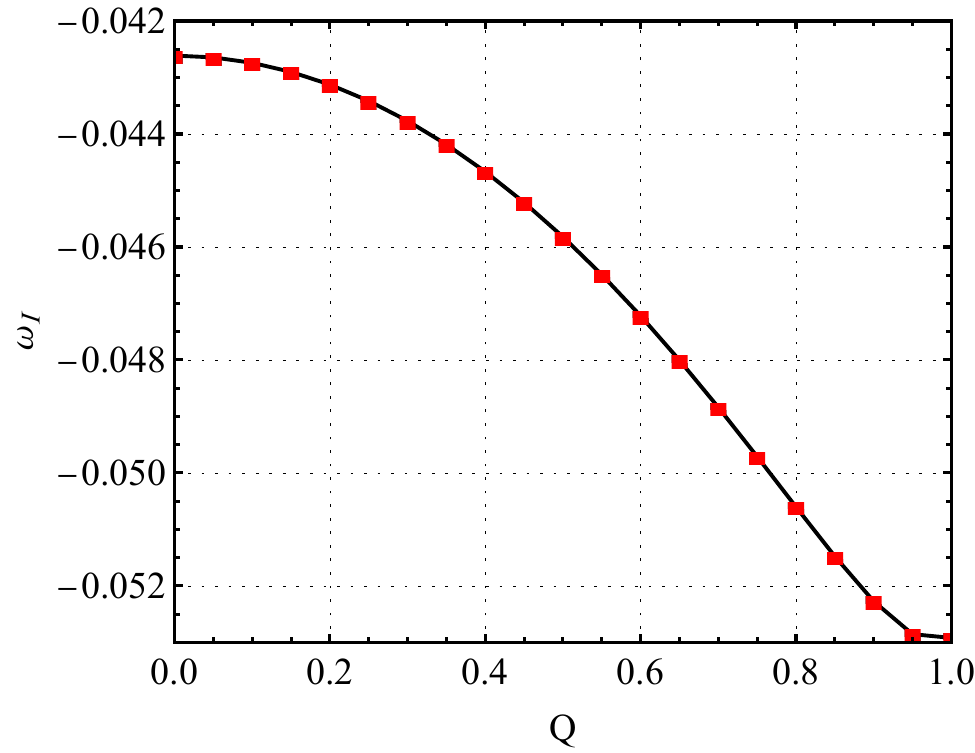}} \vspace{-0.2cm}
		\caption{{\color{black}Variation of quasinormal modes with $M=1$, $n= 0, l=3, N_s= -0.1, \beta = 0.1$ and $\Lambda = 0.02$ for massless scalar
				perturbation. }}
		\label{qnm03}
	\end{figure}

	To have a better realization of the quasinormal mode dependency on the model parameters of the black hole, we have plotted the real and imaginary quasinormal modes with respect to the model parameters. In Fig. \ref{qnm01}, we have shown the variation of real and imaginary quasinormal modes with respect to the Rastall parameter $\beta$. One can see that $\beta$ impacts the oscillation frequencies of ring-down GWs non-linearly. With an increase in $\beta$, oscillation frequency decreases drastically. The decay or damping rate of GWs also decreases non-linearly with an increase in the value of $\beta$. It seems that the violation of energy-momentum conservation can have significant impacts on ring-down GWs.
	
	In Fig. \ref{qnm02}, we have shown the variation of quasinormal modes with respect to the black hole structural parameter $N_s$. It is important to the fact that the value of $N_s$ is closely associated with weak energy conditions. For positive values of $N_s$, the weak energy condition is violated. In such a scenario, both the oscillation frequency of ring - down GWs or real quasinormal modes and the damping rate increase significantly. 
	
	Finally, in Fig. \ref{qnm03}, we have shown the impacts of black hole charge on the quasinormal modes. It is seen that an increase in $Q$ increases the real part of quasinormal modes non-linearly. In the case of the damping rate also, we observe a non-linear increase of the damping rate with an increase in the value of $Q$. However, near $Q = 1$, a slight reverse pattern is observed. 
	
	{\color{black}
		Our investigation predicts that the surrounding field (quintessence, in our case) can have significant impacts on the quasinormal mode spectrum of a black hole in Rastall gravity. It is important in the framework of Rastall gravity because the theory deviates from GR only in the presence of a non-vanishing energy-momentum tensor. Moreover, black hole charge also has a noticeable impact on the quasinormal mode spectrum. It is interesting to note that the impact of energy-momentum conservation violation has an opposite impact on the quasinormal mode spectrum in comparison to both charge and quintessence fields.
	}
	
	{\color{black} In a broader sense, the insights acquired from these investigations are crucial for comprehending the complex physics of black holes. They show non-trivial correlations between model parameters like the Rastall parameter, black hole structural parameters, and charge, as well as the behavior of quasinormal modes. Such knowledge helps our understanding of the peculiar dynamics and properties of black holes, particularly within the context of Rastall gravity, and lays the groundwork for future study on this intriguing topic.}
	
	Our investigation also suggests that recent observational constraints on Rastall gravity \cite{Tang} can have significant impacts on the quasinormal mode spectrum. In the near future, utilizing observational results from LISA, we can have a constraint on the theory from GWs and then it might be possible to check whether constraints put by Ref. \cite{Tang} stands in agreement with this.
	
	\section{Evolution of Scalar Perturbations on the Black hole Geometry}\label{sec7}
	
	In the preceding section of our investigation, our primary focus was on the numerical computation of the quasinormal modes and their behavior about the model parameters such as black hole charge, structural constant, and Rastall parameter. Now, our attention turns to the examination of the time domain profiles of scalar perturbations. This analysis necessitates the utilization of a highly effective technique called the time domain integration formalism, which was originally introduced by Gundlach \cite{gundlach}.
	
	To commence our exploration, we establish the scalar field as $\psi(r^*,
	t) = \psi(i \Delta r^*, j \Delta t) = \psi_{i,j} $, where $r^*$ represents the tortoise coordinate and $t$ denotes time. Correspondingly, we define the potential as $V(r(r^*)) = V(r^*,t) =
	V_{i,j}$. By incorporating these definitions, we can express the governing equation of the scalar field in a discretized form:
	\begin{equation}
		\dfrac{\psi_{i+1,j} - 2\psi_{i,j} + \psi_{i-1,j}}{\Delta r^{*2}} - \dfrac{
			\psi_{i,j+1} - 2\psi_{i,j} + \psi_{i,j-1}}{\Delta t^2} - V_i\psi_{i,j} = 0.
	\end{equation}
	
	To initiate the temporal evolution of the scalar field, we establish the following initial conditions: $\psi(r^*,t) = \exp \left[ -\dfrac{(r^*-k_1)^2%
	}{2\sigma^2} \right]$, representing a Gaussian wave-packet, and $\psi(r^*,t)\vert_{t<0} = 0$. In this context, $k_1$ and $\sigma$ correspond to the median and width of the initial wave packet, respectively.
	
	By employing an iterative scheme, we can compute the time evolution of the scalar field employing as follows:
	
	\begin{equation}
		\psi_{i,j+1} = -\,\psi_{i, j-1} + \left( \dfrac{\Delta t}{\Delta r^*}
		\right)^2 (\psi_{i+1, j + \psi_{i-1, j}}) + \left( 2-2\left( \dfrac{\Delta t%
		}{\Delta r^*} \right)^2 - V_i \Delta t^2 \right) \psi_{i,j}.
	\end{equation}
	
	To obtain the profile of $\psi$ with respect to time $t$, we apply the aforementioned iterative scheme. It is crucial to note that throughout the numerical procedure, we must select a fixed value for the ratio $\frac{\Delta t}{\Delta r^*}$ and ensure that it remains less than 1. This restriction guarantees the satisfaction of the Von Neumann stability condition, thereby preserving the numerical stability of our calculations.
	
	Through the implementation of the time domain integration formalism described above, we can gain valuable insights into the temporal characteristics of scalar perturbations. This enables us to obtain a comprehensive understanding of the dynamic behavior exhibited by the system under investigation.

	\begin{figure}[htbp]
		\centerline{
			\includegraphics[scale = 0.8]{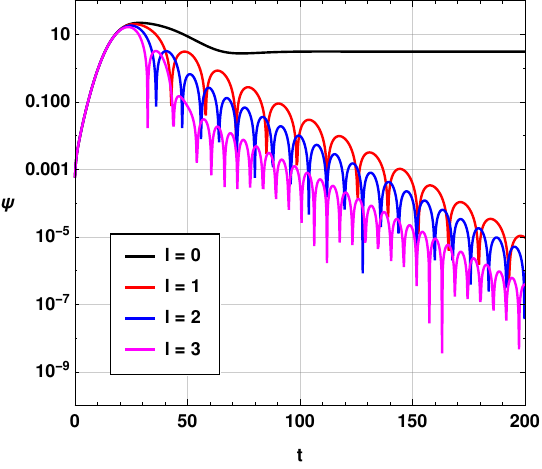}\hspace{0.5cm}
			\includegraphics[scale = 0.8]{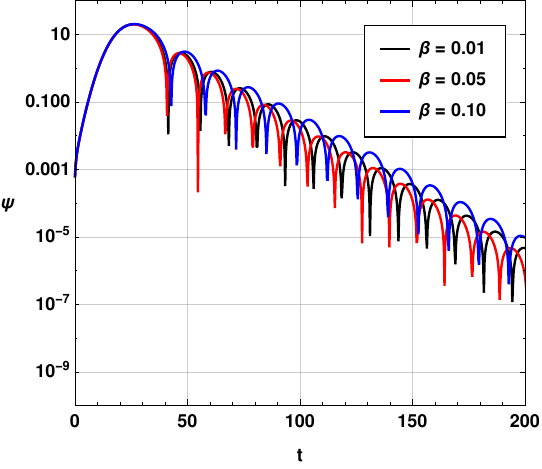}} \vspace{-0.2cm}
		\caption{ Time domain profiles with $M=1$, $n= 0, Q = 0.3$ and $\Lambda = 0.02$ for massless scalar
			perturbation. {\color{black} We have used  $\beta = 0.1$ and $ N_s= -0.01$ on the first panel and  $l=1$ and $N_s= -0.01$ on the second panel.} }
		\label{time01}
	\end{figure}
	
	\begin{figure}[htbp]
		\centerline{
			\includegraphics[scale = 0.8]{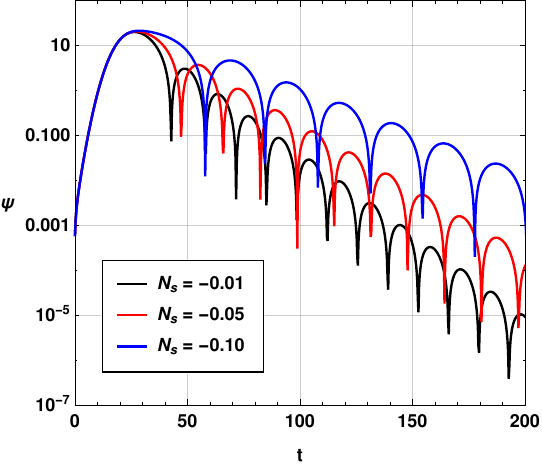}\hspace{0.5cm}
			\includegraphics[scale = 0.8]{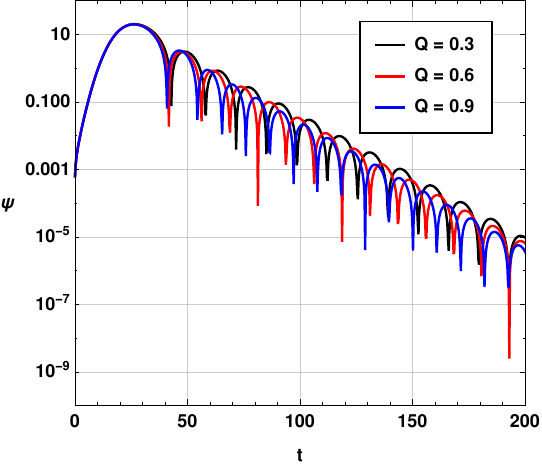}} \vspace{-0.2cm}
		\caption{Time domain profiles with $M=1$, $n= 0$, $l=1$, $\beta = 0.1$ and $\Lambda = 0.02$ for massless scalar
			perturbation. {\color{black} We have used  $Q = 0.3$ on the first panel and  $N_s= -0.01$ on the second panel.} }
		\label{time02}
	\end{figure}
	
	We have shown the time evolution profiles for massless scalar perturbation for different values of multipole moment $l$ on the first panel of Fig. \ref{time01}. One can see that with an increase in the value of multipole moment $l$, the decay rate or damping rate as well as the oscillation frequency of ring-down GWs increases. For $l=0$, one may note that the perturbation is unstable and corresponds to a positive value of imaginary quasinormal mode as seen from the previous results.
	
	The impact of energy-momentum conservation violation is shown on the second panel of Fig. \ref{time01}. The time domain profiles depict a significant impact of the Rastall parameter $\beta$ on the time domain profiles of scalar perturbation. Similarly, we have shown the impacts of black hole structural parameter $N_s$ and charge $Q$ on the time domain profiles on the first and second panels of Fig. \ref{time02} respectively. These results show that the black hole hairs or the model parameters can have different impacts on the time domain profiles and quasinormal mode spectrum. In the near future, with observational results from space-based GW detectors like LISA, it might be possible to constrain such theories effectively. Moreover, the detection of ring-down GWs can also shed some light on the possibility of energy-momentum conservation violation.
	
	\section{Scattering and Greybody factor}\label{sec8}
\begin{figure*}[t!]
      	\centering{
      	\includegraphics[scale=0.35]{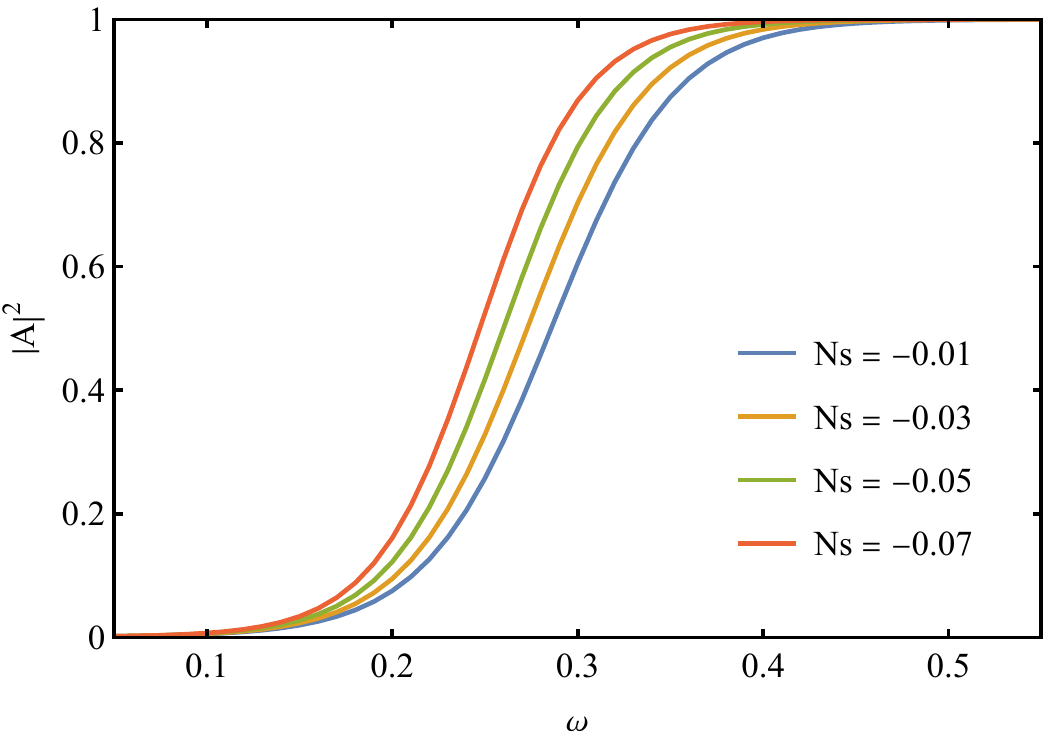}
       \includegraphics[scale=0.35]{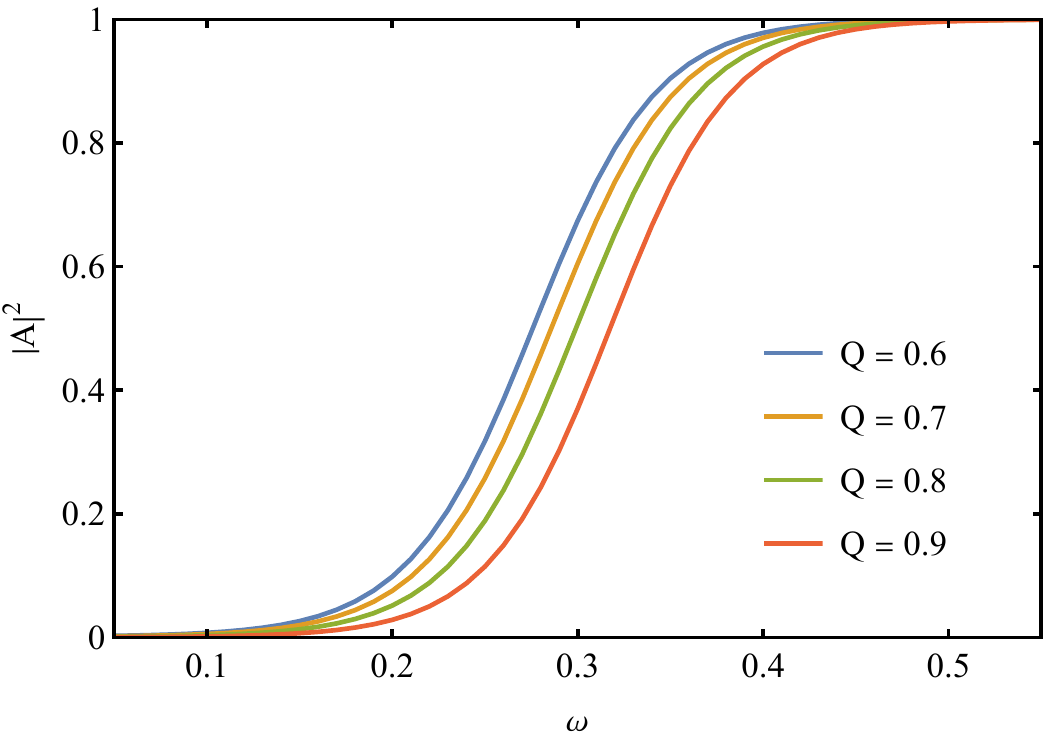}
       \includegraphics[scale=0.35]{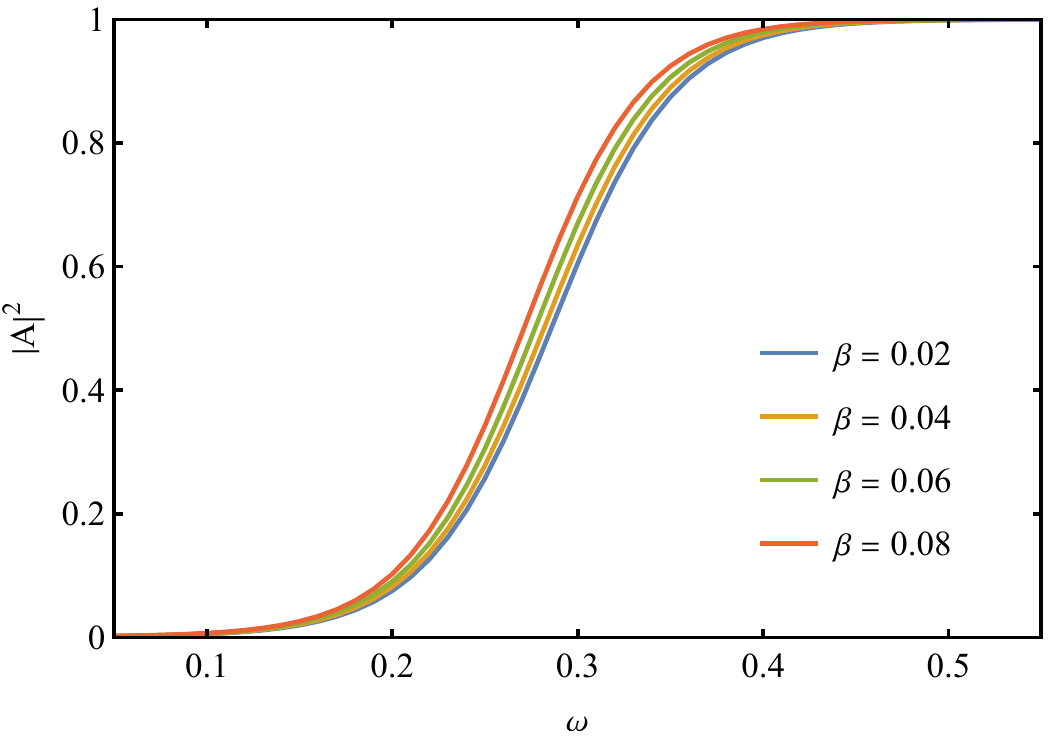}}
      	\caption{ {\color{black} The greybody factors of the scalar field are computed using the third-order WKB method for $M=1$ and $l=1$. The first panel shows the results for $Q=0.7$, $\beta=0.02$ and $N_s=-0.01,-0.03,-0.05$ and $-0.07$, while the second panel represents the results for $N_s=-0.01$, $\beta=0.02$ and $Q=0.6,0.7,0.8$ and $0.9$ and the third panel shows the greybody factor for constant $Q=0.7$, $N_s=-0.01$ and variation of $\beta$. } }
		\label{fig:Grey}
      \end{figure*}

	In this section, we investigate the scattering process using the WKB method. The absorption cross-section is another
	an important aspect of gravitational perturbations around a black hole spacetime. the probability for an outgoing wave to reach infinity
	or the probability for an incoming wave to be absorbed by the black hole is defined as greybody factor \cite{konoplya3,cardoso, dey} which plays an important role in studying the tunneling probability of the field through the effective potential of the
	given black hole spacetime. We have shown the effective potential in Fig. \ref{fig:Veff}, demonstrating the effect of $N_s$ and $Q$ on the $V_{eff}$. Therefore, the influence of these parameters on the greybody
	factor and absorption cross-section are in our interest.\\
	Scattering via the WKB method requires boundary conditions as the fields near the horizon and at infinity are expected to have these
	asymptotic form \cite{cardoso2,zhidenko2003}
	\begin{equation}
		R_{\omega l}=
		\begin{cases}
			{{e^{ - i\omega {r^*}}} + R{e^{i\omega {r^*}}}} \quad if\ {r^*} \to -\infty ~ (r \to {r_h})\\
			{T{e^{ - i\omega {r^*}}}} \quad \quad \quad \quad if \ {r^*} \to +\infty ~ (r \to \infty )\\
		\end{cases}
	\end{equation}
	where $R$ and $T$ are the reflection and transmission coefficients, respectively. These coefficients can be obtained by
	\begin{equation}
		|R|^{2} =  \frac{1}{1+e^{-2i\pi \mathcal{K}}},
	\end{equation}
	\begin{equation}\label{Trans}
		|T|^{2} = \frac{1}{1+e^{+2i\pi \mathcal{K}}}=1-|R|^{2}
	\end{equation}
	$\mathcal{K}$ is a parameter which can be determined by following equation {\color{black} \cite{Kono2003}}
	\begin{equation}
		\mathcal{K}= \frac{i({\omega}^{2}-V_{0})}{\sqrt{-2 V''_{0}}} - \sum_{j=2}^{6} \Omega_{j}
	\end{equation}
	here $V_0$ is the maximum of the effective potential, $V''_0$ is the
	the second derivative of the effective potential in its maximum with respect to the tortoise coordinate $r^*$, and $\Omega$ are higher order WKB
	corrections which depend on derivatives of the
	effective potential at its maximum\cite{shutz,devi,Iyer}.\\
	Now, the grey body factor for each multipole number is calculated. We represent the greybody factor of the scalar field with respect to $\omega$. 
	The left panel of Fig.\ref{fig:Grey}, (a), illustrates that increasing the value of $N_s$ leads to a decrease in the grey-body factors, indicating a smaller fraction of the scalar field is penetrating the potential barrier. The right panel of Fig.\ref{fig:Grey}, (b), shows increasing the value of $Q$ has the same impact on the greybody factor as increasing $N_s$. Therefore these results are consistent with the investigation of effective potential in Fig.\ref{fig:Veff} that when for a constant $Q$ value, the $N_s$ increases or in the situation of constant $N_s$, $Q$ increases, the height of the effective potential barrier goes higher which means that a smaller fraction of the scalar field is penetrating the potential barrier or the probability for an outgoing wave to reach infinity called greybody factor decreased. On the other hand, when the value of $\beta$ goes higher from $0.04$ to $0.16$, the greybody factor experiences a higher value and shifts leftward.
	
	{\color{black}The transmission coefficient, defined in Eq. \eqref{Trans} can be used to determine the partial absorption cross section \cite{crispino2009scattering, anacleto2020absorption}.

		\begin{equation}
			{\sigma^l_{\mathrm{abs}}} = \frac{{\pi (2l + 1)}}{{{{\omega}^2}}}{\left| {{\mathrm{T}_{l}}({\omega} )} \right|^2},
		\end{equation}
		where ${\omega}$ is the frequency and $l$ is the mode number.\\
		
		We investigate the impact of different parameters on the absorption cross section in Fig. \ref{fig:Cross}. Here the first panel indicates that in the same frequency and other fixed parameters, increasing the value of $Q$ decreases the absorption cross section.
		To see the effect of $N_s$ parameter, second plot of Fig.\ref{fig:Cross} demonstrates that higher values of the absolute value of $N_s$ lead to higher absorption cross section for the same frequency and Finally, increasing $\beta$ corresponds to higher absorption cross section.
		These observations	aligns with the fact that when the the height of the effective potential barrier in Fig.\ref{fig:Veff} increases the absorption cross section goes down. For instance, in third panel of Fig.\ref{fig:Veff}, by bigger values of Rastall parameter $\beta$, the height of the potential barrier is decreasing, therefore we expect a lower absorption cross section which is consistent with the result in third panel of Fig.\ref{fig:Cross}.
		
\begin{figure*}[t!]
      	\centering{
      	\includegraphics[scale=0.35]{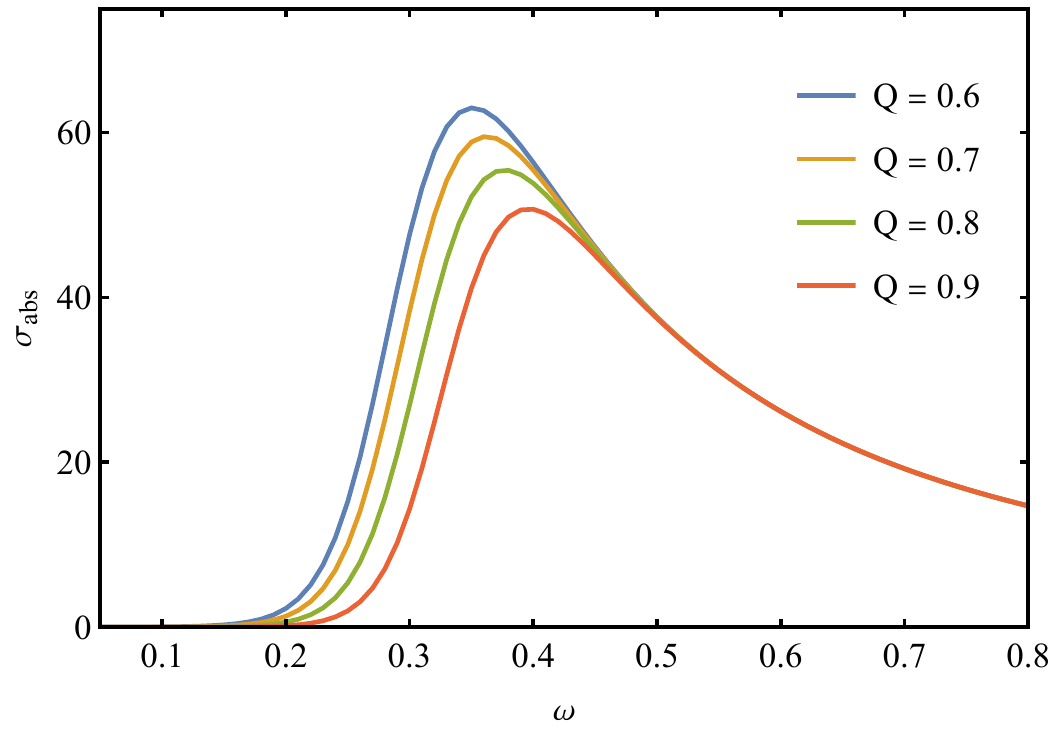}
       \includegraphics[scale=0.35]{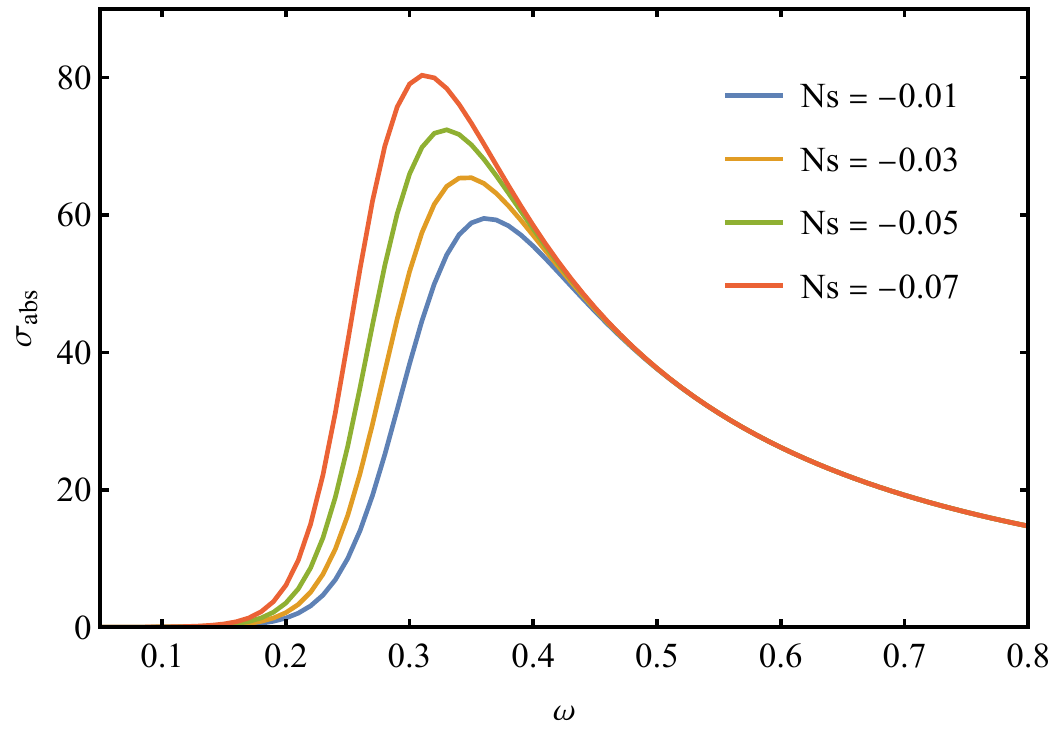}
       \includegraphics[scale=0.35]{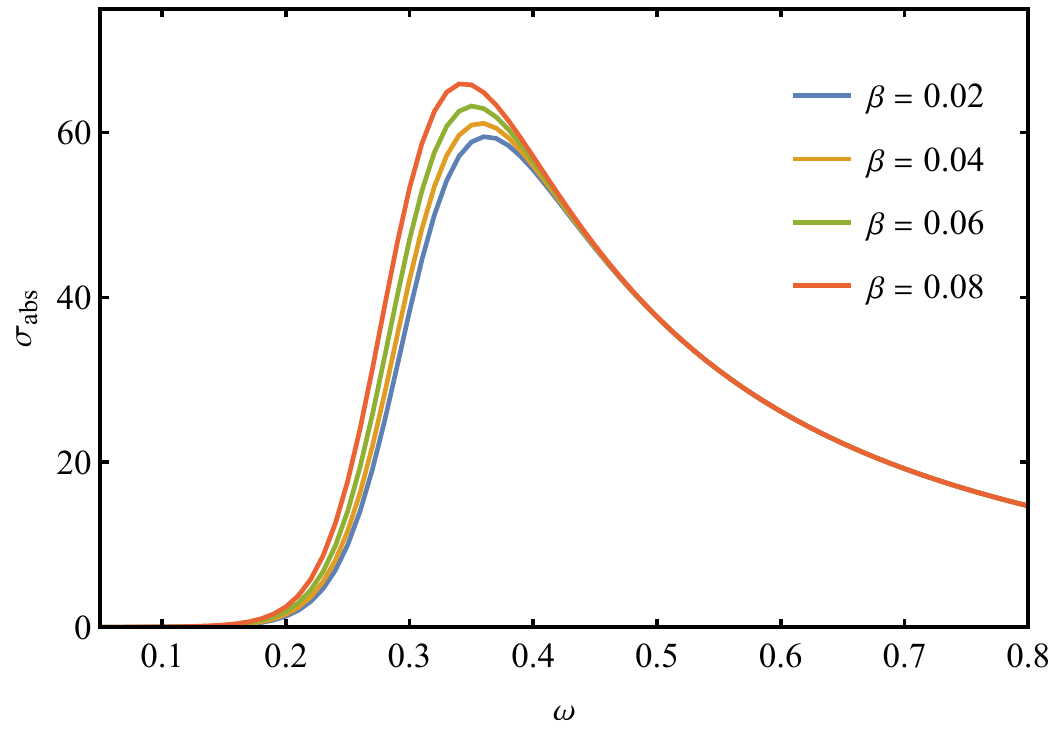}}
      	\caption{{\color{black}The partial absorption cross section of the scalar field with $l=1$ are computed for $M=1$ and $l=1$. The first panel shows the results for $N_s=-0.01$, $\beta=0.02$ and $Q=0.6,0.7,0.8$ and $0.9$, while the second plot represents the results for $Q=0.7$, $\beta=0.02$ and $N_s=-0.01,-0.03,-0.05$ and $ -0.07$ and the third panel represents the absorption cross section for constant $Q=0.7$, $N_s=-0.01$ and variation of $\beta$. }}
			\label{fig:Cross}
      \end{figure*}

		
		\section{Geodesics} \label{sec9}

		The trajectory of a light ray can be deflected due to the presence of a gravitational field, which is a consequence of general relativity. It has gained lots of attention with the recently revealed image of a black hole by the Event Horizon Telescope. Understanding the properties of astrophysical black holes requires an investigation of geodesics.  provide.
		
		By exploring the behavior of light in the vicinity of the black hole, we could gain insight into the black hole parameters. We would like to investigate the null geodesics equation in the spacetime that corresponds to Rastall gravity and analyze the effect of the Rastall parameter on light trajectory. We consider the spherically symmetric spacetime metric given by Eq \eqref{metric01} and apply the Lagrangian defined as.
		
		\begin{equation}\label{Lgr}
			\mathcal{L} =  - \frac{1}{2}( - f(r){(\frac{{dt}}{{d\lambda }})^2} + \frac{1}{{f(r)}}{(\frac{{dr}}{{d\lambda }})^2} + {r^2}{(\frac{{d\theta }}{{d\lambda }})^2} + {r^2}{{\mathop{\rm Sin}\nolimits} ^2}\theta {(\frac{{d\varphi }}{{d\lambda }})^2}
		\end{equation}
		
		where $\lambda$ is the arbitrary affine parameter. As there are two Killing vector $\partial_t$ and $\partial_r$, two constants of motion can be defined as $E$ and $L$ with the following equations
		\begin{align}\label{cons1}
			&f(r){{\dot t}} = E\\ \label{cons2}
			&{r^2}{{\mathop{\rm Sin}\nolimits} }\theta {{\dot \varphi }} = L
		\end{align}
		
		In this work, the angle $\theta$ is considered $\pi/2$ and $\dot{\theta}=0$ as initial conditions. It means that we only work on the equatorial plane without loss of generality. 
		Now we apply Eq. \eqref{cons1} and \eqref{cons2}, the Lagrangian in Eq.\eqref{Lgr} could be written as
		\begin{equation}
			{{\dot r}^2} + f(r)(\frac{{{L^2}}}{{{r^2}}} + \varepsilon ) - E^2 = 0
		\end{equation}
		Here, $\epsilon$ is equal to $2L$, $\epsilon=0$ and $\epsilon=1$ are associated with massless and massive particle, respectively \cite{fernando}. In this study, we examine the light which corresponds to $\epsilon0+=0$. Now, we  rewritten the above equation as
		\begin{equation}\label{rdot}
			{\left(\frac{{\mathrm{d}r}}{{\mathrm{d}\lambda }}\right)^2} + {\tilde{V}_{eff}}(r) = 0.
		\end{equation}
		where the effective potential is 
		\begin{equation}\label{veffr}
			{\tilde{V}_{eff}}(r) = {f(r) }(\frac{L^2}{{{r^2}}} -\frac{E^2}{f(r)})
		\end{equation}
		By proposing a new variable $u=1/r$, and an impact parameter $b=\frac{L}{E}$, we rewrite the Eq. \eqref{rdot} in the following form
		\begin{equation}\label{geoline}
			\frac{{du}}{{d\varphi }} =\sqrt{ \frac{1}{{{b^2}}} - {u^2}f(u)}=\sqrt{\frac{1}{b^2}+\frac{\Lambda }{3-12 \beta }+2 M u^3-N_s u^{2-\frac{4 \beta }{1-2 \beta }}-Q^2 u^4-u^2}
		\end{equation}
		By considering the differentiation of the above equation, we solve Eq.\eqref{geoline} numerically which leads to the equation of light trajectory in the equatorial plane.
		
\begin{figure*}[t!]
      	\centering{
      	\includegraphics[scale=0.35]{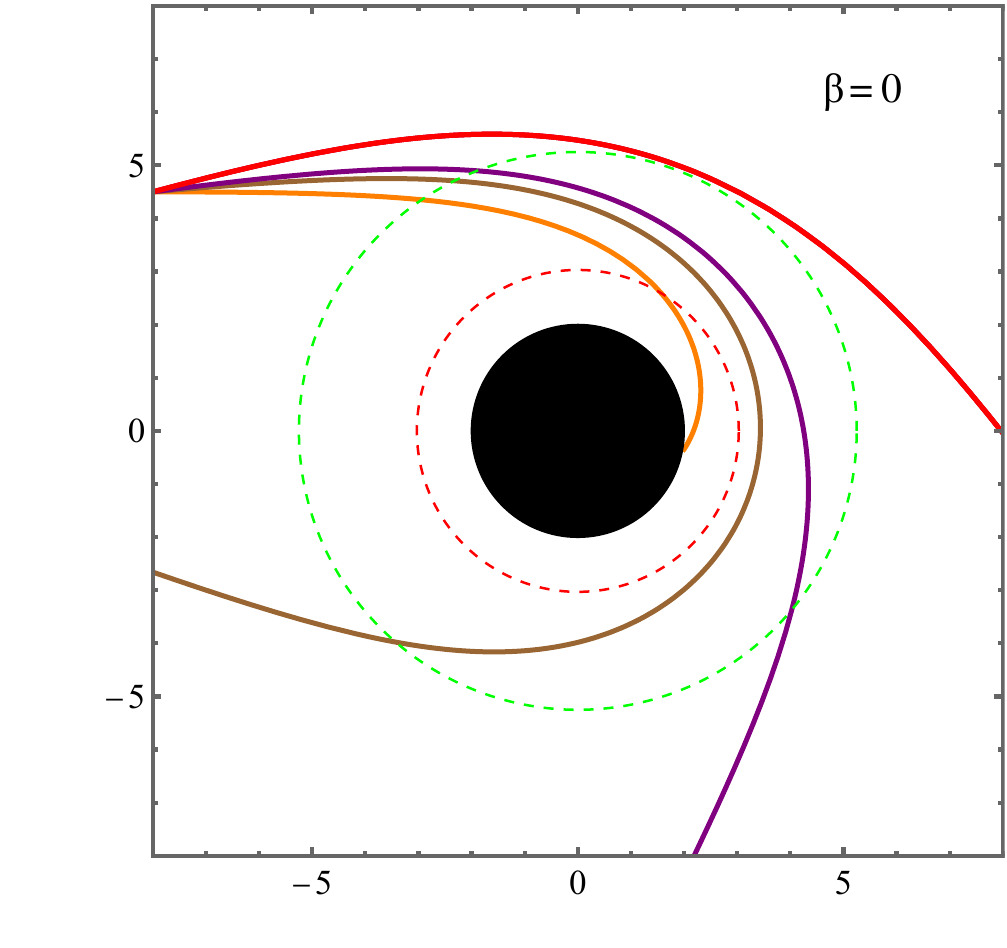}
       \includegraphics[scale=0.35]{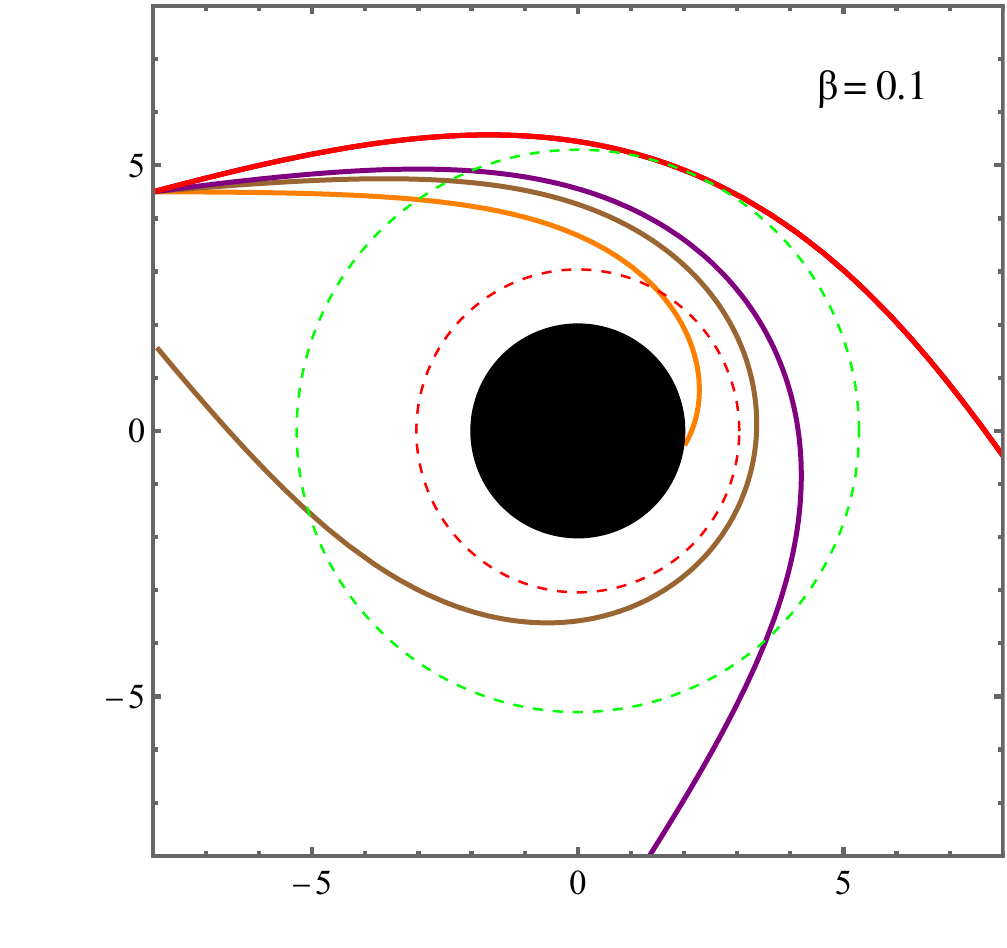}
       \includegraphics[scale=0.35]{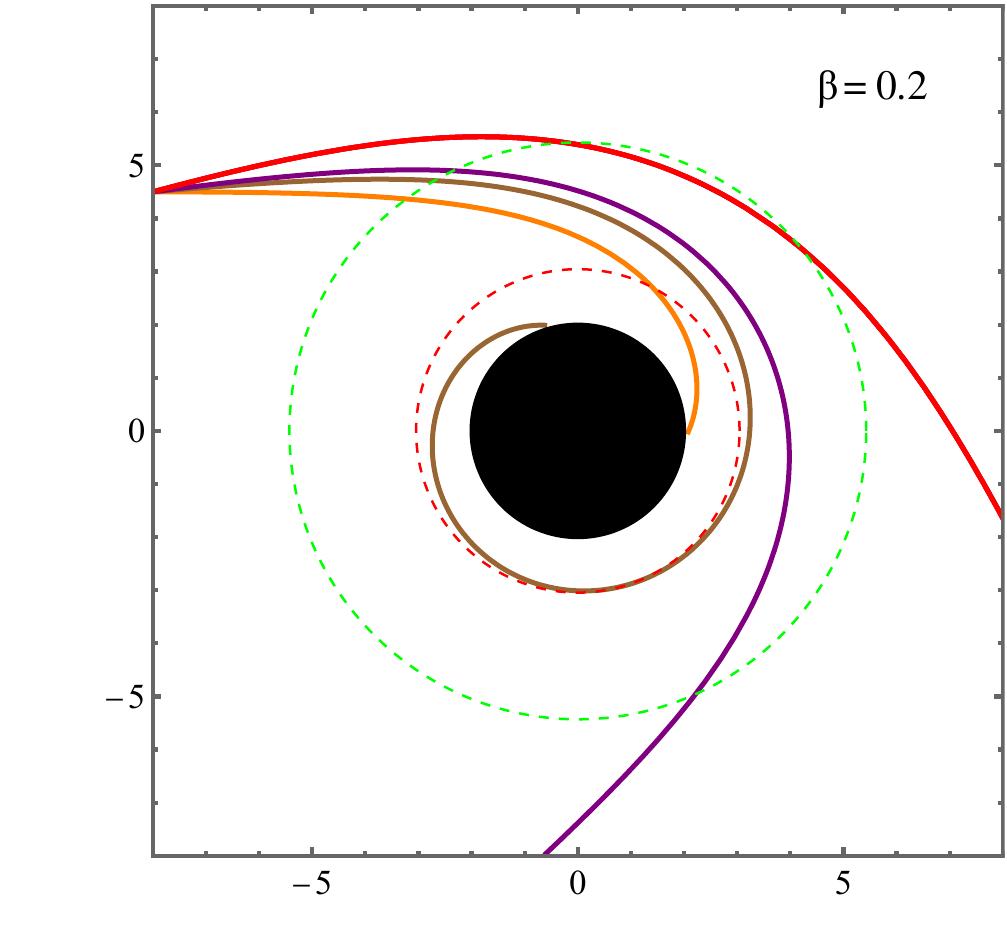}}
      	\caption{{\color{black} The light trajectory for $M=1$, $Q=0.01$, $N_s=-0.01$ and $\Lambda=0.02$ with (a) $\beta=0$, (b) $\beta=0.1$ and (c) $\beta=0.2$}}
			\label{fig:trajectory}
      \end{figure*}

		Furthermore, The photonic radius ($r_{ph}$) is found by considering the unstable condition on effective potential as 
		\begin{equation}
			{\tilde{V}_{eff}} = \frac{{\mathrm{d}{\tilde{V}_{eff}}}}{{\mathrm{d}r}} = 0,
		\end{equation}
		Therefore the following equation is obtained for the photonic radius
		\begin{align}\label{rps}
			&2 {f(r_{ph})}- {r_{ph}f'(r_{ph})} = 0,\\
			&-3 M r_{ph}+\frac{(4 \beta -1) N_s r_{ph}^{\frac{2}{1-2 \beta }}}{2 \beta -1}+2 Q^2+r_{ph}^2=0
		\end{align}
		On the other hand, the shadow radius is calculated using celestial coordinates proposed as $X = \mathop {\lim }\limits_{{r_0} \to \infty } ( - {r_0}^2{\mathop{\rm Sin}\nolimits} {\theta _0}\frac{{d\varphi }}{{dr}}{|_{{r_0},{\theta _0}}})$ and $Y = \mathop {\lim }\limits_{{r_0} \to \infty } ({r_0}^2\frac{{d\theta }}{{dr}}{|_{{r_0},{\theta _0}}})$, where $r_0$ and $\theta_0$ denotes the position of the observer at infinity. The shadow radius can be expressed as 
		\begin{equation}
			{R_{Shadow}} = \sqrt {{X ^2} + {Y ^2}} = \frac{{{r_{ph}}}}{{\sqrt {f({r_{ph}})} }}= \frac{{{r_{ph}}}}{{\sqrt {1-\frac{2 M}{{r_{ph}}}+\frac{Q^2}{{r_{ph}}^2}+N_s {r_{ph}}^{\frac{4 \beta }{1-2 \beta }}-\frac{\Lambda {r_{ph}}^2}{3-12 \beta }} }},
		\end{equation}
		Note that ${r_{ph}}$ is the radius of the photon sphere calculated in Eq. \ref{rps}.
		
		In Ref.\cite{vagnoz}, two limits for shadow are represented based on
		the EHT horizon-scale image of $Sgr A$, after averaging the Keck and VLTI mass-to-distance ratio priors for 
		\begin{equation}
			4.55 < {R_{Shadow}} < 5.22
		\end{equation}
		and 
		\begin{equation}
			4.21 < {R_{Shadow}} < 5.56
		\end{equation}
		
		we have tried to find a boundary for a characteristic parameter according to experimental data. For instance, when $M = 1$, $N_s = -0.01$, $\beta = 0.002$ the shadow radius versus $Q$ is plotted and the above mentioned experimental constraints are represented by two pairs of horizontal lines in Fig. \ref{fig:cons}.
		\begin{figure}[h]
			\centering
			\includegraphics[scale = 0.35]{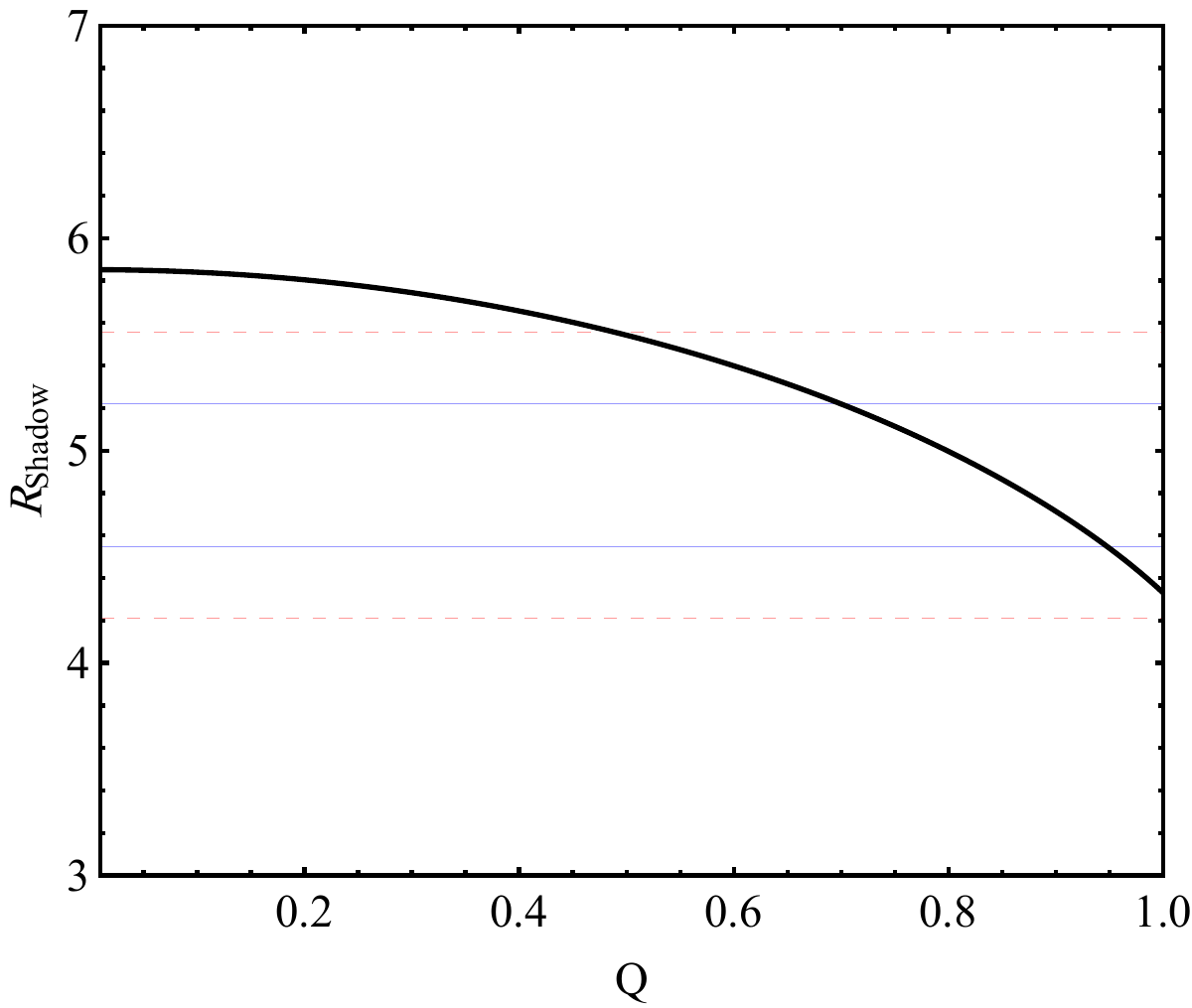}
			\caption{{\color{black} Shadow radius versus $Q$ for $M=1$, $N_s= -0.01$ , $\beta = 0.002$, $\Lambda = 0.02$, the experimental constraints are shown as horizontal lines in $4.21$, $5.56$ and $4.55$ , $5.22$}}
			\label{fig:cons}
		\end{figure}
		
		We can see that for the fixed parameter we have chosen, the obtained constraint for $Q$ is $ 0.69 < Q < 0.94 $ for the first experimental constraint and $0.48 < Q < 1$
		for the second experimental boundary.\\
		The behavior of the geodesic lines for different values of the Rastall parameter are shown in Fig. \ref{fig:trajectory}. The red and green dashed circles indicate the photonic and shadow radius, respectively. We plot the light trajectory for three different values of the Rastall parameter. In each plot, four different impact parameters are indicated with different colors.
		According to these plots, it is obvious that a higher value of $\beta$ empowers the gravitational effect of the black hole on the light. For instance, if we compare the brown line in plots (a), (b), and (c), when $\beta$ is zero, the light turns around the black hole, in the next plot when $\beta$ is $0.1$ the light line bends more and cross the vertical axes in upper value. Finally, when $\beta$ equals $0.2$, the light is trapped by the black hole's gravitation and cannot escape like the same condition light in plots (a) and (b).
		
	}

	\section{Concluding Remarks}
	\label{sec10}
	
	In this work, {\color{black} first we have discussed the metric function and Hawking temperature. The Rastall parameter influenced the remnant mass, noting that the higher Rastall parameter yielded a smaller remnant radius as well as remnant mass. Additionally,} we have investigated the quasinormal modes, scattering, and greybody factors of AdS/dS Reissner-Nordstr\"om black hole surrounded by quintessence field in Rastall gravity. We have found that the black hole structural constant as well as charge have noticeable impacts on the quasinormal mode spectrum of the black hole.
	It is observed that for $l=0$ and $n=0$, one can have unstable modes with positive imaginary quasinormal modes corresponding to a smaller value of black hole charge $Q$. However, for a higher black hole charge, the black hole spacetime becomes stable emitting physical quasinormal modes.
	It is seen from the Eq. \eqref{rho} that for smaller values of $\beta$, a positive structural constant $N_s$ results in violation of the weak energy condition which can be avoided by considering large values of $\beta$ or negative values of $\beta$. Since large values of $\beta$ may give rise to unstable quasinormal modes, we chose negative $N_s$ values to get weak energy conditions respecting scenario. Our investigation shows that $N_s$ has a linear impact on the quasinormal modes of the black hole and for negative $N_s$ values, both real quasinormal modes and damping rates are smaller.
	Rastall parameter $\beta$ and black hole charge $Q$ are found to have non-linear impacts on the quasinormal modes. 
	
	In the case of greybody factors, we observe that both $N_s$ and $Q$ have similar types of impacts on the greybody factors. By the conditions mentioned for Fig.\ref{fig:Grey}, increasing $Q$ for the constant condition of other values leads to a higher greybody factor and also a higher absolute value of $N_s$ with other constant parameters results in increasing of greybody factor, but $\beta$ has opposite influence and when $\beta$ increases, greybody factor goes down.
	{\color{black} Moreover, the absorption cross section has been obtained highlighting that when the $Q$ goes high the absorption cross section decreases, despite the effect of $\beta$ that bigger values of $\beta$ made higher absorption cross section at the same frequency. We observed the same influence for a higher value of the absolute value of $N_s$. Additionally, we explored the geodesics of massless particles. The Rastall parameter has a significant impact on the light trajectory and higher values of $\beta$ made the gravitational properties of black holes stronger and the deflection of light increased with this parameter.}	
	
	In conclusion, our investigation of the quasinormal modes, scattering, and greybody factors of the dS Reissner-Nordstr\" om black hole surrounded by a quintessence field in Rastall gravity has revealed the significant impacts of the black hole structural constant and charge on the quasinormal mode spectrum, emphasizing stability and physical emission. We have also found that the Rastall parameter and black hole charge have non-linear effects on the quasinormal modes, while the black hole structural constant and charge exhibit similar trends in their influence on the greybody factors. Incorporating recent observational constraints on Rastall gravity \cite{Tang} and future data from LISA, we anticipate obtaining further constraints on the theory through GWs, allowing us to assess the agreement between the imposed constraints and emerging evidence, thereby enhancing our understanding of Rastall gravity's compatibility with observational data.
	

\end{document}